\documentclass[12pt,%thmsa,%
notitlepage,twoside,a4paper]{article}%
\usepackage{amssymb}
\usepackage{amsmath}
\usepackage{amsfonts}%
\usepackage{graphicx}
\begin{document}

\begin{center}
{\Large Mixing and Coherent Structures in 2D Viscous Flows }

\markright\today\smallskip by

{\large H.W. Capel}

Institute of Theoretical Physics, Univ. of Amsterdam

Valckenierstraat 65, 1018 XE Amsterdam, The Netherlands

and

{\large R.A. Pasmanter}\footnote{Corresponding author. E-mail:
pasmante@knmi.nl}

KNMI, P.O.Box 201, 3730 AE De Bilt, The Netherlands

\medskip Abstract
\end{center}

\noindent We introduce a dynamical description based on a probability density
$\phi(\sigma,x,y,t)$ of the vorticity $\sigma$ in two-dimensional viscous
flows such that the average vorticity evolves according to the Navier-Stokes
equations. A time-dependent mixing index is defined and the class of
probability densities that maximizes this index is studied. The time
dependence of the Lagrange multipliers can be chosen in such a way that the
masses $m(\sigma,t):=\int\!dxdy\,\phi(\sigma,x,y,t)$ associated with each
vorticity value $\sigma$ are conserved. When the masses $m(\sigma,t)$ are
conserved then 1) the mixing index satisfies an H-theorem and 2) the mixing
index is the time-dependent analogue of the entropy employed in the
statistical mechanical theory of inviscid 2D flows [Miller, Weichman \& Cross,
Phys. Rev. A \textbf{45} (1992); Robert \& Sommeria, Phys. Rev. Lett.
\textbf{69}, 2776 (1992)]. Within this framework we also show how to
reconstruct the probability density of the quasi-stationary coherent
structures from the experimentally determined vorticity-stream function
relations and we provide a connection between this probability density and an
appropriate initial distribution.

\section{ Introduction}

When studying the dynamics of two-dimensional fluid motion characterized by a
vorticity field $\omega(x,y,t)$ it can be useful to turn to a statistical
description with probability distributions $\phi(\sigma,x,y,t)$ for the
microscopic vorticity $\sigma$ such the average value of $\sigma$ over these
distributions is equal to $\omega(x,y,t).$ In particular, this has been done
in the description of the quasi-stationary states (QSS), i.e., the coherent
structures which are often reached in (numerical) experiments after a fast
mixing process has taken place\cite{mcwilliams,flor,montPFA2,marteau,segre}.
At high Reynolds' numbers, the vorticity fields $\omega_{S}(x,y)$ of these
QSS's satisfy $\omega-\psi$ relations to a good approximation, i.e.,
$\omega_{S}(x,y)\simeq\Omega(\psi(x,y))$ where $\psi(x,y)$ is the
corresponding stream-function. In other words, the QSS's are approximate
stationary solutions of the Euler equation.

More specifically, in the early 1990's Miller\cite{millerPRL} and
Robert\cite{robertfrans,robertJSP} together with their
coworkers\cite{millerPRA,robsom,robrelax}presented a statistical mechanical
theory of steady flows in \emph{inviscid}, two-dimensional fluids, an approach
that can be traced back to Linden-Bell's work of 1967 \cite{lyndenBell}. Some
outstanding aspects of this non-dissipative system are: 1) an infinite number
of conserved quantities, the masses associated with each microscopic-vorticity
value $\sigma,$ 2) non-uniform steady states or coherent structures, 3)
negative-temperature states (already predicted by Onsager's work on point
vortices \cite{onsager} ). Theoretical predictions were compared with
numerical simulations and with experimental measurements in quasi-two
dimensional fluids, e.g., in
\cite{matthaeSelect,marteau,brands,segre,branChav}. However, under standard
laboratory conditions fluids are viscous. Similarly, numerical simulations
require the introduction of a non-vanishing (hyper)viscosity in order to avoid
some numerical instabilities and other artifacts. A non-vanishing viscosity
can lead to noticeable effects, like the breakdown of the conservation laws,
even at high Reynolds' numbers, i.e., when the viscosity \textquotedblleft is
small\textquotedblright. This is true especially when studying long-time
processes like the generation of the coherent structures. In spite of the
dissipative nature of the flows studied, in many cases, it was found that the
agreement between the theoretical predictions based on the Miller, Robert and
Sommeria (MRS) inviscid theory and (numerical) experiments was better than expected.

The main purpose of the present paper is to better understand these issues in
a more dynamical setting. We start by reviewing the MRS theory in Section
\ref{MRS}. In Section \ref{MicroModels} we consider \emph{viscous} flows and
propose a family of model evolution equations for the vorticity distribution
$\phi(\sigma,x,y,t).$ In Sections \ref{Mixing}--\ref{mixingSectn} we discuss a
class of time-dependent distributions that maximize a mixing index under
certain constraints. In particular, in Section \ref{fourCharacterztn} it is
shown that the time-dependent Lagrange multipliers appearing in these
distributions can be chosen in such a way that the masses associated with each
microscopic-vorticity value $\sigma$ are conserved. When these masses are
conserved, the mixing index is the time-dependent analogue of the entropy
functional used by MRS and it satisfies an H-theorem, as it is shown in
Appendix A. The distribution $\widetilde{\phi}_{S}(\sigma,\psi(x,y))$
associated with a given QSS can be obtained, at least in principle, by
addressing the reconstruction problem, i.e., the question of how to extract
its defining parameters from the QSS's $\omega-\psi$ relation. This is
discussed in Section \ref{reconstructing}.\ In doing so we provide a natural
framework for a time-dependent statistical theory connecting an appropriate
initial distribution to the QSS distribution associated with the experimental
$\omega-\psi$ relation and evolving in agreement with the Navier-Stokes
equation. The relation of the present work with the MRS theory and with the
yardsticks' conditions of reference \cite{Capel} is discussed in Section
\ref{conclusions}.

\section{The Miller-Robert-Sommeria (MRS) Theory\label{MRS}}

\subsection{Review}

The pillars on which the statistical mechanical
approach\cite{millerPRA,robsom} stands are the conserved quantities of the
non-dissipative Euler equations. These quantities are: 1) the total energy
$E_{o}:=2^{-1}\int_{A}\!dx\,dy\,v_{o}^{2}$ where $\overrightarrow{v}_{o}(x,y)$
is the initial velocity field and $A$ is the area occupied by the fluid and 2)
the area density, denoted by $g(\sigma)\,d\sigma,$ occupied by fluid with
vorticity values between $\sigma$ and $\sigma+d\sigma,$ which is%
\[
g(\sigma):=\int_{A}\!dx\,dy\,\phi(\sigma,x,y),
\]
where $\phi(\sigma,x,y)\geq0$\ is the probability density of finding a
microscopic vorticity value $\sigma$ at position $(x,y),$ with%
\begin{equation}
\int\!d\sigma\,\phi(\sigma,x,y)=1,\;\mathrm{for}\;\mathrm{all}%
\;(x,y).\label{uno}%
\end{equation}
For the sake of simplicity, we will ignore other conserved quantities which
may be present due to some symmetries of the domain $A$, like the linear
momentum and/or the angular momentum. Averages taken over this distribution
will be indicated by $\left\langle \cdots\right\rangle ,$ i.e.,
\begin{equation}
\left\langle f(\sigma,x,y)\right\rangle :=\int\!d\sigma\,f(\sigma
,x,y)\phi(\sigma,x,y).\label{averages}%
\end{equation}
The macroscopic vorticity $\omega(x,y)$\ is then%
\[
\omega(x,y)=\left\langle \sigma\right\rangle .
\]
It is assumed that the initial area density, denoted by $g_{o}(\sigma),$ is
given by%
\begin{equation}
g_{o}(\sigma):=\int_{A}\!dx\,dy\,\delta(\sigma-\omega_{o}(x,y)),\label{Go}%
\end{equation}
where $\omega_{o}(x,y)=\nabla\times\overrightarrow{v_{o}}$ is the initial
macroscopic vorticity field. \newline As usual, one derives the probability
distribution for observing, on a microscopic level, a vorticity value $\sigma$
by maximizing the entropy $S$ under the constraints defined by the conserved
quantities. The entropy used by Miller, Robert and Sommeria (MRS) is
\begin{equation}
S:=-A^{-1}\int_{A}\!dx\,dy\,\int\!d\sigma\,\phi(\sigma,x,y)\ln\phi
(\sigma,x,y),\label{entropy}%
\end{equation}
The probability distribution $\phi_{S}(\sigma,x,y)$ that one obtains by
maximizing the entropy $S$ is
\begin{align}
\phi_{S}(\sigma,x,y) &  :=Z^{-1}\exp\left[  -\beta\sigma\psi(x,y)+\mu
(\sigma)\right]  ,\label{w-psi}\\
\mathrm{with\;}Z(\psi(x,y)) &  :=\int\!d\sigma\,\exp\left[  -\beta\sigma
\psi(x,y)+\mu(\sigma)\right]  ,\nonumber
\end{align}
where $\psi(x,y)$ is the stream function, i.e., $-\Delta\psi(x,y)=\omega
_{S}(x,y),$ with $\omega_{S}(x,y)$ the macroscopic vorticity field in the most
probable state, i.e.,
\[
\omega_{S}(x,y)=\int\!d\sigma\,\sigma\phi_{S}(\sigma,x,y).
\]
We have then the $\omega$--$\psi$ relation%
\begin{align*}
\omega_{S}(x,y)  & =\Omega(\psi(x,y)),\\
\mathrm{with}\ \ \ \ \Omega(\psi)  & :=\int\!d\sigma\,\sigma\widetilde{\phi
}_{S}(\sigma,\psi),\\
\mathrm{where}\ \ \ \ \widetilde{\phi}_{S}(\sigma,\psi(x,y))  & :=\phi
_{S}(\sigma,x,y),
\end{align*}
which, in an experimental context, is often called the scatter-plot. This is a
mean-field approximation, valid for this system \cite{millerPRA}, see also
eq.\thinspace(\ref{meanfield}) below.\ In eq.\thinspace(\ref{w-psi}), $\beta$
and $\mu(\sigma)$ are Lagrange multipliers such that the QSS energy $E_{S}$
\[
E_{S}=\frac{1}{2}\int_{A}\!dx\,dy\,\omega_{S}(x,y)\psi(x,y)
\]
and the QSS microscopic-vorticity area distribution $g_{S}(\sigma),$%
\begin{equation}
g_{S}(\sigma):=\int_{A}\!dx\,dy\,\phi_{S}(\sigma,x,y),\label{g_mrs}%
\end{equation}
have the same values as in the initial vorticity field, i.e., $E_{S}=E_{o}$
and $g_{S}(\sigma)=g_{o}(\sigma).$ The system of equations is closed by
\begin{align}
\omega_{S}(x,y) &  =-\Delta\psi,\label{meanfield}\\
\mathrm{i.e.},\;\ \ \ \Omega(\psi) &  =-\Delta\psi.
\end{align}
which embodies the mean-field approximation.

Since by construction $g_{S}(\sigma)=g_{o}(\sigma),$ the assumption expressed
by Eq. (\ref{Go}) implies that $M_{n}$\ the \emph{microscopic} vorticity
moments in the quasi-stationary state,%
\begin{equation}
M_{n}:=\int\!d\sigma\,\sigma^{n}g_{S}(\sigma),\label{momentsM}%
\end{equation}
equal the \emph{macroscopic} vorticity moments in the initial state,
\[
\int\!d\sigma\,\sigma^{n}g_{o}(\sigma)=\int_{A}\!dx\,dy\,\omega_{o}^{n}(x,y),
\]
confer equation (\ref{Go}).

\subsection{The initial state\label{initial state}}

As we have seen, in its standard formulation, the MRS theory introduces an
asymmetry in the microscopic characterization of the initial state and that of
the final quasi-stationary state. Namely, it is \emph{assumed} that the
initial state is an \textquotedblleft unmixed state\textquotedblright\ for
which the microscopic and macroscopic description coincide. This allows to
determine $g_{o}(\sigma)$ the \emph{microscopic} vorticity distribution of the
initial state from the \emph{macroscopic} vorticity field $\omega_{o}(x,y),$
confer eq.\thinspace(\ref{Go}). By contrast, the microscopic vorticity-area
density of the most probable state, $g_{S}(\sigma),$ is given by
$\int\!dx\,dy\,\phi_{S}(\sigma,x,y),$ confer eq.\thinspace(\ref{g_mrs}). In
most cases, $G_{S}(\sigma),$ the \emph{macroscopic} vorticity density of the
most probable state,
\begin{equation}
G_{S}(\sigma):=\int_{A}\!dx\,dy\,\delta(\sigma-\omega_{S}(x,y)),\label{Gs}%
\end{equation}
is different from the \emph{microscopic} vorticity-area density of the most
probable state, i.e., $G_{S}(\sigma)\neq g_{S}(\sigma).$ For example, for
non-vanishing viscosity the even moments of $G_{S}(\sigma)$ denoted by
$\Gamma_{2n},$ i.e.,%
\[
\Gamma_{2n}:=\int_{A}\!dx\,dy\,\omega_{S}^{2n}(x,y),
\]
\ are smaller than those of $g_{S}(\sigma)$ since in the MRS approach one
assumes that $g_{S}(\sigma)=g_{o}(\sigma),$ confer (\ref{decay}) below. This
difference between the moments of the macroscopic and those of the microscopic
vorticity has led to confusion and to discussions in the literature related to
the interpretation of the MRS theory\cite{come,respo}. \newline According to
the MRS theory the microscopic-vorticity moments $M_{n}$ of the QSS, confer
Equation (\ref{momentsM}), equal the microscopic-vorticity moments of the
$\delta$-type initial distribution defined in (\ref{Go}). In Subsection III B
of an earlier paper\cite{Capel} we have expressed this infinite set of
equalities in terms of the $\omega-\psi$ relation and the initial macroscopic
vorticity field $\omega_{o}(x,y).$ In order to assess the validity of the MRS
approach, we introduced then a yardstick associated with each moment $M_{n},$
see further the discussion in Section \ref{conclusions}.

\section{Microscopic Viscous Models\label{MicroModels}}

From here onwards we deal with time dependent probability densities, i.e.,
$\phi(\sigma,x,y,t)d\sigma$ is the probability of finding at time $t$ a
microscopic vorticity value in the range $(\sigma,\sigma+d\sigma)$ at a
position $(x,y)$ which is non-negative and normalized
\begin{equation}
\int\!d\sigma\,\phi(\sigma,x,y,t)=1.\label{normalztn}%
\end{equation}
The macroscopic vorticity field is
\begin{equation}
\omega(x,y,t)=\left\langle \sigma\right\rangle ,\label{vortDef}%
\end{equation}
where the notation introduced in (\ref{averages}) has been extended to the
time-dependent case. In the inviscid case, the time evolution of $\phi
(\sigma,x,y,t)$ can be taken to be
\begin{equation}
\frac{\partial\phi(\sigma,x,y,t)}{\partial t}+\vec{v}(x,y,t)\cdot\nabla
\phi(\sigma,x,y,t)=0,\label{zeroth}%
\end{equation}
where the macroscopic, incompressible velocity field $\vec{v}(x,y,t)$
satisfies appropriate boundary conditions and is related to the macroscopic
vorticity $\omega(x,y,t)$ by%
\begin{equation}
\nabla\times\vec{v}=\omega\widetilde{z},\label{rotor}%
\end{equation}
with $\widetilde{z}$ a unit vector perpendicular to the $(x,y)$-plane.
Consequently, the advective term in equation (\ref{zeroth}) is quadratic in
$\phi$ and different values of $\sigma$ are coupled by this term.

In this Section we introduce some model evolution equations for the
probability density $\phi(\sigma,x,y,t).$ We demand that these microscopic
models be compatible with the macroscopic Navier-Stokes equations. The general
form of these models is
\begin{equation}
\frac{\partial\phi(\sigma,x,y,t)}{\partial t}+\vec{v}(x,y,t)\cdot\nabla
\phi(\sigma,x,y,t)=\nu O(\sigma,x,y,t),\label{defO}%
\end{equation}
with $\nu$ the fluid viscosity and $O$ as yet undefined but constrained by 1)
the conservation of the total probabilty $\int\!d\sigma\,\phi(\sigma,x,y,t),$
therefore,
\begin{equation}
\int\!d\sigma\,O=0,\label{zeroth2}%
\end{equation}
and by 2) the macroscopic Navier-Stokes equation should follow from the
microscopic model, therefore,%
\begin{equation}
\int\!d\sigma\,\sigma O=\Delta\omega(x,y,t),\label{sigmaO}%
\end{equation}
so that, multiplying both sides of equation (\ref{defO}) by $\sigma,$
integrating them over $\sigma$ and making use of\thinspace(\ref{vortDef}) and
(\ref{sigmaO}), one gets the Navier-Stokes equation
\begin{equation}
\frac{\partial\omega(x,y,t)}{\partial t}+\vec{v}(x,y,t)\cdot\nabla
\omega(x,y,t)=\nu\Delta\omega(x,y,t).\label{NS}%
\end{equation}
For future convenience, let us introduce
\[
O=:\Delta\phi(\sigma,x,y,t)+\overline{O}\phi,
\]
with
\begin{subequations}
\begin{align}
\left\langle \overline{O}\right\rangle  & =0,\label{Ozero}\\
\mathrm{and}\ \ \left\langle \sigma\overline{O}\right\rangle  & =0,\nonumber
\end{align}
so that the constraints (\ref{zeroth2}) and (\ref{sigmaO}) are satisfied.

It is convenient to introduce the \textquotedblleft masses\textquotedblright%
\ $m(\sigma,t)$\ associated with each value $\sigma$ of the microscopic
vorticity,
\end{subequations}
\begin{equation}
m(\sigma,t):=\int\!dx\,dy\,\phi(\sigma,x,y,t).\label{charges}%
\end{equation}
In the inviscid case, incompressibility and the fact that the vorticity is
just advected by the velocity field, allow for a complete association between
a vorticity value $\sigma$ and the area that is occupied by such value. In
this case one talks of `conservation of the area occupied by a vorticity
value'. In fact, in the inviscid case, $\nu=0,$ equation (\ref{defO}) has a
solution $\phi(\sigma,x,y,t)=\delta(\sigma-\omega(x,y,t))$ so that the
above-defined $m(\sigma,t)$ is indeed the area occupied by the vorticity field
with value $\sigma.$ As soon as we introduce a diffusion process, as it is
implied by equation\thinspace(\ref{defO}) with $\nu\neq0,$ such an
identification becomes problematic if not impossible. It is for this reason
that we call $m(\sigma,t)$ a \textquotedblleft mass\textquotedblright\ and, by
doing so, we stress the obvious analogy with an advection-diffusion process of
an infinite number of \textquotedblleft chemical species\textquotedblright,
one species for each value $\sigma.$

The time derivative of these masses is%
\begin{equation}
\frac{\partial m(\sigma,t)}{\partial t}=\nu\int\!dx\,dy\,\overline{O}%
\phi(\sigma,x,y,t).\label{conservMass}%
\end{equation}
In order to derive these time-evolution equations (\ref{conservMass}) it was
assumed that there is no leakage of $\phi(\sigma,x,y,t)$ through the boundary,
i.e., that the total flux of $\phi(\sigma,x,y,t)$ through the boundary
vanishes,
\[
\oint ds\,\vec{n}\cdot\left[  \vec{v}(x,y,t)\phi(\sigma,x,y,t)-\nu\vec{\nabla
}\phi(\sigma,x,y,t)\right]  =0,
\]
where the path of the integral is taken over the boundary and $\vec{n}$ is the
normal unit vector. These conditions, as well as other boundary conditions
that will be used in the sequel, are satisfied, e.g., in the case of periodic
boundary conditions as well as by probability fields whose support stays away
from the boundary at all times. In the case of impenetrable boundary
conditions one has that, on the boundary, $\,\vec{n}\cdot\vec{v}\equiv0,$ so
that the last condition reduces to
\begin{equation}
\oint ds\,\vec{n}\cdot\vec{\nabla}\phi(\sigma,x,y,t)=0.
\end{equation}
On a macroscopic level this implies, e.g., that
\[
\oint ds\,\vec{n}\cdot\vec{\nabla}\omega(x,y,t)=0
\]
and the conservation of the total circulation $\int\!dx\,dy\,\omega(x,y,t). $

\label{Model 1}The simplest model satisfying the above requirements is the one
with $\overline{O}\equiv0,$ i.e.,
\begin{equation}
\frac{\partial\phi(\sigma,x,y,t)}{\partial t}+\vec{v}(x,y,t)\cdot\nabla
\phi(\sigma,x,y,t)=\nu\Delta\phi(\sigma,x,y,t),\label{simple}%
\end{equation}
with the incompressible velocity field related to the vorticity as in
eq.\thinspace(\ref{rotor}). In spite of its simplicity, this model is very
instructive because, while it dissipates energy, it has an infinite number of
conserved quantities. Indeed, the masses are conserved,
\begin{equation}
\frac{\partial m(\sigma,t)}{\partial t}=0,\label{conserva}%
\end{equation}
confer (\ref{conservMass}).

One of the consequences of the conservation laws (\ref{conserva}) is that all
the microscopic-vorticity moments $M_{n}(t):=\int\!d\sigma\,\sigma^{n}%
m(\sigma,t)$ are constants of the motion, i.e.,
\[
\frac{dM_{n}}{dt}=0.
\]
In the sequel we shall assume that the conservation of all the microscopic
moments $M_{n}$ implies in turn that the masses $m(\sigma,t)$\ are conserved.
This is the case if certain technical conditions are satisfied, see e.g.,
\cite{shohat}. \newline On the other hand, all even moments of the
\textbf{macroscopic} vorticity, call them $\Gamma_{2n}(t),$ i.e., all the
quantities
\[
\Gamma_{2n}(t):=\int\!dx\,dy\,\omega^{2n}(x,y,t),
\]
in particular the \textbf{macroscopic} enstrophy $\int\!dx\,dy\,\omega
^{2}(x,y,t),$ are dissipated,
\begin{equation}
\frac{d\Gamma_{2n}}{dt}=-\nu2n(2n-1)\int\!dx\,dy\,\omega^{2(n-1)}\left\vert
\nabla\omega\right\vert ^{2}\leq0,\label{decay}%
\end{equation}
as it is implied by the Navier-Stokes equation (\ref{NS}). Also the energy
$E=1/2\int\,dxdy\,v^{2}$ is dissipated
\[
\frac{dE}{dt}=\nu\int\!dx\,dy\,\vec{v}\cdot\Delta\vec{v}=-\nu\int
\!dx\,dy\left[  \,\left(  \frac{\partial^{2}\psi}{\partial x^{2}}\right)
^{2}+\left(  \frac{\partial^{2}\psi}{\partial y^{2}}\right)  ^{2}+2\left(
\frac{\partial^{2}\psi}{\partial x\partial y}\right)  ^{2}\right]  \leq0,
\]
where $\psi(x,y,t)$ is the stream-function associated with $\vec{v}%
(x,y,t)$\ and appropriate boundary conditions, e.g., periodic boundary
conditions, have been assumed. Under these boundary conditions, one also has,%
\begin{align*}
E  & =\frac{1}{2}\int\!dx\,dy\,\omega\psi,\\
\mathrm{and}\ \ \ \frac{dE}{dt}  & =-\nu\int\!dx\,dy\,\omega^{2}.
\end{align*}
Notice that the microscopic energy and the macroscopic one coincide,
\[
-\frac{1}{2}\int_{A}\!dx\,dy\,\int\!d\sigma\,\int_{A}\!dx^{\prime}%
\,dy^{\prime}\,\int\!d\sigma^{\prime}\,\sigma\phi(\sigma,x,y,t)K(x,y,x^{\prime
},y^{\prime})\sigma^{\prime}\phi(\sigma^{\prime},x^{\prime},y^{\prime},t)=
\]%
\[
-\frac{1}{2}\int_{A}\!dx\,dy\,\int_{A}\!dx^{\prime}\,dy^{\prime}%
\,\omega(x,y,t)K(x,y,x^{\prime},y^{\prime})\omega(x^{\prime},y^{\prime},t)
\]
In this last expression, $K(x,y,x^{\prime},y^{\prime})$ is the Green function
solving
\[
\Delta K(x,y,x^{\prime},y^{\prime})=\delta(x-x^{\prime})\delta(y-y^{\prime}),
\]
and the appropriate boundary conditions.

\emph{In conclusion: The viscous Navier-Stokes equation (\ref{NS}) does not
exclude the possibility of an infinite number of conserved quantities
}$m(\sigma,t)$\emph{\ as defined in equation (\ref{charges}).}

In the following Sections we will consider more generic models with
$\overline{O}\neq0$ that conserve the masses $m(\sigma,t).$

\section{Chaotic mixing\label{Mixing}}

Considering once more the models defined by equations (\ref{simple}) and
(\ref{defO}), we notice that they are advection-diffusion equations for a
non-passive scalar $\phi$ with the viscosity $\nu$ playing the role of a
diffusion coefficient. This type of equations has been studied extensively,
see e.g. \cite{ottino} and the references therein. It is well-known that a
time-dependent velocity field $\vec{v}(x,y,t)$ usually leads to chaotic
trajectories, i.e., to the explosive growth of small-scale $\phi$-gradients.
These small-scale gradients are then rapidly smoothed out by diffusion, the
net result being a very large effective diffusion coefficient, large in
comparison to the molecular coefficient $\nu.$

Based on these observations we will consider situations such that during a
period of time, starting at $t_{0}$ and ending at $T_{S}$ when a
quasi-stationary structure (QSS) is formed, a fast mixing process takes place
leading to a probability density, denoted by $\phi_{S}\left(  \sigma
,x,y\right)  ,$ that maximizes the spatial spreading or mixing of the masses
$m(\sigma,T_{S})$ and is such that $\omega_{S}(x,y)=\int\!d\sigma\,\sigma
\phi_{S}(\sigma,x,y)$ satisfies the experimentally found $\omega-\psi$
relation. More specifically, we will investigate solutions of equation
(\ref{defO}) for suitable $\nu O$ on the right-hand side satisfying the
following conditions: i) at time $t=T_{S}$ the solution $\phi(\sigma,x,y,t)$
is the above-mentioned maximally mixed $\phi_{S}(\sigma,x,y)$ complying with
the experimental $\omega-\psi$ relation and ii) the mixing takes place much
faster than the changes in the masses $m(\sigma,t),$ so that $m(\sigma
,T_{S})\cong m(\sigma,t_{0}).$\ In order to express these ideas in a
quantitative form, we need a mathematical definition of the degree of
spreading or mixing of a solute's mass in a domain. As described in Appendix
A, the $0$--order degree of mixing $s_{0}(\sigma,t)$ defined as%
\begin{equation}
s_{0}(\sigma,t):=-A^{-1}\int\!dx\,dy\,\phi(\sigma,x,y,t)\ln\left[
A\phi(\sigma,x,y,t)/m(\sigma,t)\right]  ,\label{so}%
\end{equation}
is the right quantity in order to quantify the mixing of the
microscopic-vorticity masses $m(\sigma,t).$ Accordingly, the total vorticity
mixing at time $t,$ is measured by%
\[
S_{0}(t):=\int\!d\sigma\,s_{0}(\sigma,t).
\]
More explicitly,%
\begin{equation}
S_{0}(t)=-A^{-1}\int\!d\sigma\int\!dx\,dy\,\phi(\sigma,x,y,t)\ln\phi
(\sigma,x,y,t)+A^{-1}\int\!d\sigma\,m(\sigma,t)\ln\left[  A^{-1}%
m(\sigma,t)\right]  .\label{largeSo}%
\end{equation}
The first term coincides with the MRS entropy (\ref{entropy}). If the masses
are conserved, i.e., if $m(\sigma,t)=m\left(  \sigma,t_{0}\right)  ,$ then the
second term is constant in time. This allows us to give a mathematical
definition of fast mixing, namely: fast mixing of $m(\sigma,t)$ takes place in
a time interval $(t_{0},T_{S})$ if the following inequalities hold,
\begin{equation}
\left\vert \frac{\partial s_{0}(\sigma,t)}{\partial t}\right\vert \gg\frac
{1}{A}\left\vert \frac{\partial m(\sigma,t)}{\partial t}\right\vert
,\ \ \mathrm{for}\ \ t_{0}\leq t\leq T_{S}.\label{muchFaster}%
\end{equation}
Accordingly, the corresponding microscopic vorticity probability density
$\phi_{S}(\sigma,x,y)$ maximizes the total degree of mixing $S_{0}(T_{S}%
)=\int\!d\sigma\,s_{0}(\sigma,T_{S})$ under the constraint that the change in
the masses $m(\sigma,t)$ during the time interval $[t_{o},T_{S}]$ is very small.

It turns out that, for our purposes, a second constraint is needed. Here we
discuss two possible choices of this second constraint. One possible choice of
the second constraint consists in using the macroscopic vorticity
$\omega(x,y,t),$ i.e., the solution of the Navier-Stokes equation with initial
condition $\omega_{o}(x,y)$ as input. This means that one demands,%
\[
\int\!d\sigma\,\sigma\phi(\sigma,x,y,t)=\omega(x,y,t).
\]
Two points should be stressed: 1) since $\omega(x,y,t)$\ is a solution of the
Navier-Stokes equation (\ref{NS}) this choice of the second constraint is
totally compatible with the presence of viscous dissipation and 2) this choice
of the second constraint can be imposed at all times $t,$ not only at time
$T_{S}$ when the QSS is present. The probability distributions which are
obtained by maximizing the total degree of mixing $S_{0}(t)$ under the
constraints of given values $m(\sigma,t)$ for the masses (\ref{charges}) and
given the vorticity field $\omega(x,y,t),$ are presented in the next Section.

The second possible choice of the constraint stems from physical evidence
that, in high Reynolds' number, two-dimensional flows, the energy is
transported from the small to the large scales and, consequently, it is weakly
affected by viscous dissipation. More formally,%
\begin{equation}
\frac{dS_{0}(t)}{dt}\gg\frac{1}{E}\left\vert \frac{dE}{dt}\right\vert
,\ \ \mathrm{for}\ \ t_{0}\leq t\leq T_{S},\label{fasterE}%
\end{equation}
Therefore, one maximazes $S_{0}$\ under the constraints that the masses and
the energy at time $T_{S}$ have some given values $m(\sigma,T_{S})$ and
$E(T_{S}).$ Introducing the corresponding Lagrange multipliers $\beta$ and
$\widetilde{\mu}(\sigma),$ as well as a Lagrange multiplier $\gamma(x,y)$
associated with the normalization constraint (\ref{normalztn}), the
constrained variation of the degree of mixing \ $S_{0}$ leads to
\begin{align}
0 &  =\beta\sigma\psi(x,y)+\widetilde{\mu}(\sigma)+\gamma(x,y)+\ln\frac
{A\phi_{S}(\sigma,x,y)}{m\left(  \sigma\right)  },\label{energico}\\
\mathrm{i.e.,\ \ \ \ }\;\phi_{S}(\sigma,x,y) &  =A^{-1}m(\sigma)\exp
(-\beta\sigma\psi(x,y)-\widetilde{\mu}(\sigma)-\gamma(x,y)),\nonumber
\end{align}
where $m(\sigma):=m(\sigma,T_{S}).$ Since $A^{-1}m(\sigma)\geq0,$ we can
define $\mu(\sigma):=-\widetilde{\mu}(\sigma)+\ln A^{-1}m(\sigma)$ and
implementing the normalization constraint (\ref{uno}), one arrives at the
probability density (\ref{w-psi}) of the inviscid, statistical mechanics
approach. The new elements here are the conditions expressed by
(\ref{muchFaster}) and (\ref{fasterE}), i.e., a criterium for the
applicability of these equations in the case of \textbf{viscous} flows.
Notice, moreover, that in contraposition to the statistical mechanical
approach, we do not require that the energy $E(T_{S})$ and the masses
$m(\sigma,T_{S})$ at time $T_{S}$ be equal to their initial values; equations
(\ref{muchFaster}) and (\ref{fasterE}) only express that the changes in these
quantities are much smaller than the change in total vorticity mixing in the
time interval $(t_{0},T_{S}).$ In the presence of viscosity $E(T_{S})\leq
E(t_{o}),$ when the equality holds one recovers exactly the MRS expressions.

In closing this Section, it is worthwhile recalling that under the physical
conditions leading to the inequality in (\ref{fasterE}) the changes in energy
are usually much smaller than the changes in enstrophy. This fact is at the
basis of the so-called selective-decay hypothesis\cite{matthaeSelect}, i.e.,
the conjecture that the QSSs correspond to macroscopic vorticity fields with
energy $E(T_{S})$ which minimize the macroscopic enstrophy $\Gamma_{2}%
(T_{S}).$ See also \cite{branChav}.

\section{The time-dependent extremal distributions\label{fourCharacterztn}}

In this Section we investigate the time-dependent probability distributions
which are obtained by using the macroscopic vorticity $\omega(x,y,t)$\ as
second constraint. As it will be shown in the following Subsection, the form
of these distributions is,%
\begin{align}
\phi^{\ast}(\sigma,x,y,t)  & =Z^{-1}\exp\left[  \mu(\sigma,t)+\chi
(x,y,t)\sigma\right]  ,\label{ansatz}\\
\mathrm{with\;\ \ \ \ }Z  & :=\int\!d\sigma\,\exp\left[  \sigma\chi
(x,y,t)+\mu(\sigma,t)\right]  .
\end{align}
The functions $\chi(x,y,t)$ and $\mu(\sigma,t)$ will be called the
\textquotedblleft potentials\textquotedblright. In contrast to the situation
in the statistical mechanics approach, these potentials can be time-dependent.
Evidently, for these distributions one has $\phi^{\ast}(\sigma
,x,y,t)=:\widetilde{\phi}(\sigma,\chi(x,y,t),t),$ therefore the $(x,y)-$%
dependence of $\omega(x,y,t)$ as well as that of all the higher-order local
moments $m_{n}(x,y,t):=\left\langle \sigma^{n}\right\rangle $ and the centered
local moments $K_{n}(x,y,t):=\left\langle \left(  \sigma-\omega\right)
^{n}\right\rangle $\ is only through $\chi(x,y,t),$ i.e., $K_{n}%
(x,y,t)=:\widetilde{K}_{n}(\chi(x,y,t),t),$ moreover,%
\begin{align*}
\omega(x,y,t)  & =:\widetilde{\Omega}(\chi(x,y,t),t),\\
\mathrm{and}\ \ \widetilde{m}_{n}(\chi,t)  & :=\frac{\int\!d\sigma\,\sigma
^{n}\exp\left[  \mu(\sigma,t)+\chi\sigma\right]  }{\int\!d\sigma\,\exp\left[
\mu(\sigma,t)+\chi\sigma\right]  },\\
\mathrm{moreover}\ \ \frac{\partial\widetilde{\Omega}}{\partial\chi}  &
=\widetilde{K}_{2}(\chi,t),\ \ \mathrm{and}\ \ \widetilde{K}_{3}%
(\chi,t):=\frac{\partial\widetilde{K}_{2}(\chi,t)}{\partial\chi}.
\end{align*}
As it is easily checked, the local moments $\widetilde{m}_{n}(\chi,t)$ satisfy
then $\partial\widetilde{m}_{n}/\partial\chi=\widetilde{m}_{n+1}%
-\widetilde{\Omega}\widetilde{m}_{n},$ i.e., the following recursion relation
holds,
\begin{align}
\widetilde{m}_{n+1}(\chi,t)  & =\mathcal{L}\circ\widetilde{m}_{n}%
(\chi,t),\label{opertL}\\
\mathrm{with}\ \ \ \ \mathcal{L}  & :=\frac{\partial\ }{\partial\chi
}+\widetilde{\Omega}(\chi,t).
\end{align}

\subsection{Maximum mixing}

Time-dependent distributions as in Equation (\ref{ansatz}) are obtained by
maximizing $S_{0}(t),$ the total degree of mixing at time $t,$ under the
constraints of i) normalization, confer Equation (\ref{normalztn}), ii) given
values $m(\sigma,t)$ for the masses at time $t$ as defined in (\ref{charges})
and iii) given vorticity field $\omega(x,y,t)$\ at time $t,$ i.e., the
distribution's first moment (\ref{vortDef}), $\left\langle \sigma\right\rangle
=\omega(x,y,t).$\ Indeed, introducing the corresponding time-dependent
Lagrange multipliers $\gamma(x,y,t),\ \widetilde{\mu}(\sigma,t)$ and
$\chi(x,y,t)$ and denoting the extremal distribution by $\phi_{M}%
(\sigma,x,y,t),$ one has that the vanishing of the first-order variation reads%
\[
0=\gamma(x,y,t)+\chi(x,y,t)\sigma+\widetilde{\mu}(\sigma,t)+\ln\frac{A\phi
_{M}(\sigma,x,y,t)}{m(\sigma,t)},
\]
from which, in analogy with the derivation of the statistical mechanical
formulas in Section \ref{Mixing}, one obtains that $\phi_{M}(\sigma,x,y,t)$
has precisely the form of $\phi^{\ast}(\sigma,x,y,t)$ in (\ref{ansatz}), i.e.,
$\phi_{M}(\sigma,x,y,t)=\phi^{\ast}(\sigma,x,y,t)$ with $\mu(\sigma
,t):=-\widetilde{\mu}(\sigma,t)+\ln A^{-1}m(\sigma,t).$ The potential
functions $\chi(x,y,t)$ and $\mu(\sigma,t)$ should be determined from the two
following constraints
\begin{align}
\omega(x,y,t)  & =Z^{-1}\int\!d\sigma\,\sigma\exp\left[  \sigma\chi
(x,y,t)+\mu(\sigma,t)\right]  ,\label{mSigma}\\
\mathrm{and\;\ \ \ }m(\sigma,t)  & =\int\!dx\,dy\,Z^{-1}\exp\left[  \sigma
\chi(x,y,t)+\mu(\sigma,t)\right]  ,\nonumber
\end{align}
where the vorticity field $\omega(x,y,t)$ is not a stationary solution of the
Euler equations but a time-dependent solution of the Navier-Stokes equations.
The main differences with the maximizer $\phi_{S}(\sigma,x,y)$ of the previous
Section are that these distributions $\phi_{M}(\sigma,x,y,t)$ are
time-dependent and that now there is no energy constraint, therefore, the
Lagrange multiplier $\chi(x,y,t)$ is not a linear function of the
stream-function $\psi(x,y,t).$ Up till now, the time dependence of $\mu
(\sigma,t)$ has been arbitrary, in the next Subsection this time dependence
will be determined such that the masses $m(\sigma,t)$ are conserved.

\subsection{The time-dependence of $\mu(\sigma,t)$ and the conservation of the
total moments \label{time-dependent}}

Assume that at all times the probability density has the form given in
(\ref{ansatz}) which is a time dependent probability density such that
$S_{0}(t)$ attains its maximum value compatible with $\left\langle
\sigma\right\rangle =\omega(x,y,t)$ and given $m(\sigma,t).$ Inserting these
expressions in the Navier-Stokes equation (\ref{NS}) and making use of simple
algebraic equalities like,%
\begin{equation}
\frac{\partial\phi(\sigma,x,y,t)}{\partial t}=\left[  \left(  \sigma
-\omega\right)  \frac{\partial\chi}{\partial t}+\frac{\partial\mu(\sigma
,t)}{\partial t}-\left\langle \frac{\partial\mu(\sigma,t)}{\partial
t}\right\rangle \right]  \phi,\label{dphi/dt}%
\end{equation}
and%
\begin{align*}
\frac{\partial\omega}{\partial t} &  =\widetilde{K}_{2}(\chi,t)\frac
{\partial\chi}{\partial t}+\left\langle \left(  \sigma-\omega\right)
\frac{\partial\mu(\sigma,t)}{\partial t}\right\rangle ,\\
\nabla\phi &  =\left(  \sigma-\omega\right)  \phi\nabla\chi,\;\;\nabla
\omega=\widetilde{K}_{2}(\chi,t)\nabla\chi,\\
\nabla\widetilde{K}_{2} &  =\widetilde{K}_{3}(\chi,t)\nabla\chi,
\end{align*}
with $K_{n}:=\left\langle \left(  \sigma-\omega\right)  ^{n}\right\rangle ,$
leads to the following relationship between $\partial\chi/\partial t$ and
$\partial\mu/\partial t,$%
\begin{equation}
\frac{\partial\chi}{\partial t}+\vec{v}\cdot\nabla\chi-\nu\Delta\chi=\nu
\frac{K_{3}}{K_{2}}\left\vert \nabla\chi\right\vert ^{2}-\frac{1}{K_{2}%
}\left\langle \left(  \sigma-\omega\right)  \frac{\partial\mu(\sigma
,t)}{\partial t}\right\rangle ,
\end{equation}
Using this expression in order to eliminate $\partial\chi/\partial t$ from
(\ref{dphi/dt}), one finally arrives at Equation (\ref{defO}) with
$\overline{O}(\sigma,x,y,t)$ expressed in terms of $\chi(x,y,t)$ and of
$\partial\mu/\partial t.$ Namely, one obtains that
\begin{align}
\overline{O}(\sigma,x,y,t)  & =\nu^{-1}\left(  \frac{\partial\mu}{\partial
t}-\left\langle \frac{\partial\mu}{\partial t}\right\rangle \right)  -\nu
^{-1}\frac{\left(  \sigma-\omega\right)  }{K_{2}}\left\langle \left(
\sigma-\omega\right)  \frac{\partial\mu}{\partial t}\right\rangle +\label{o}\\
& +\left[  K_{2}+\frac{K_{3}}{K_{2}}\left(  \sigma-\omega\right)  -\left(
\sigma-\omega\right)  ^{2}\right]  \left\vert \nabla\chi\right\vert ^{2},
\end{align}
As one can check, this expression satisfies $\left\langle \overline
{O}\right\rangle =0$ and $\left\langle \sigma\overline{O}\right\rangle =0$
independently of the specific form of $\partial\mu/\partial t.$ In other
words, the $(x,y)$-dependent constraints (\ref{normalztn}) and (\ref{vortDef})
hold at all times and for all possible time-dependences of $\mu(\sigma,t).$

From (\ref{o}) it follows that the simplest viscous model of the type given in
Equation (\ref{ansatz}), with an identically vanishing $\overline{O}%
(\sigma,x,y,t),$ can only be realized under very special, and rather trivial,
conditions. Indeed, since $\overline{O}(\sigma,x,y,t)\equiv0$ has to hold for
any value of $\sigma,$ equation (\ref{o}) implies that $\partial\mu/\partial
t$ must be quadratic in $\sigma$ and that $\left\vert \nabla\chi\right\vert
^{2}$ may be time-dependent but must be $(x,y)$-independent. Therefore, the
simplest viscous model with an extremal distribution as given in
(\ref{ansatz}), $\overline{O}(\sigma,x,y,t)\equiv0 $ and satisfying equation
(\ref{o}) can hold only if $\chi(x,y,t),$ and $\omega(x,y,t)\,$are linear
functions of the space coordinates.

\label{conserved_moments}For general $\mu(\sigma,t)$ there is no conservation
of the masses $m(\sigma,t).$ However, choosing a suitable time-dependence of
$\mu(\sigma,t)$ such that%
\begin{equation}
\int\!dxdy\,\phi\overline{O}=0,\label{zero}%
\end{equation}
ensures the conservation of the masses, i.e., $m(\sigma,t)=m(\sigma,t_{o})$ at
all times $t,$ confer equation (\ref{conservMass}). Using equation (\ref{o})
the last equality is seen to imply that,%
\begin{align}
\int\left[  \frac{\partial\mu(\sigma,t)}{\partial t}-\left\langle
\frac{\partial\mu(\sigma,t)}{\partial t}\right\rangle -\frac{(\sigma-\omega
)}{K_{2}}\left\langle (\sigma-\omega)\frac{\partial\mu(\sigma,t)}{\partial
t}\right\rangle \right]  \phi(\sigma,x,y,t)\,dxdy  & =\label{dmu/dtBig}\\
-\nu\int\left[  K_{2}+\frac{K_{3}}{K_{2}}\left(  \sigma-\omega\right)
-\left(  \sigma-\omega\right)  ^{2}\right]  \left\vert \nabla\chi\right\vert
^{2}\phi(\sigma,x,y,t)\,dxdy  & =\nonumber
\end{align}

This equation is a complicated integro-differential equation for the
time-dependence of $\mu(\sigma,t),$ however, using a Taylor expansion
$\mu(\sigma,t)=\sum_{k}\mu_{k}(t)\sigma^{k},$ we can derive an infinite set of
linear differential equations for the $d\mu_{k}/dt.$\ In fact, multiply
equation (\ref{o}) by $\sigma^{n}\phi(\sigma,x,y,t),$ integrate it over
$\sigma$ and get then that:%
\[
\nu\left\langle \sigma^{n}\overline{O}\right\rangle =-\nu\left\vert \nabla
\chi\right\vert ^{2}h_{n2}+%
%TCIMACRO{\dsum \limits_{k=2}^{\infty}}%
%BeginExpansion
{\displaystyle\sum\limits_{k=2}^{\infty}}
%EndExpansion
h_{nk}\frac{d\mu_{k}}{dt}%
\]
with%
\[
h_{nk}(x,y,t):=m_{k+n}-m_{k}m_{n}-\frac{\left(  m_{n+1}-\omega m_{n}\right)
\left(  m_{k+1}-\omega m_{k}\right)  }{K_{2}},
\]
where $m_{n}$ are the local moments $m_{n}(x,y,t):=\left\langle \sigma
^{n}\right\rangle .$ The conservation of the moments $M_{n}=\int
\!dxdy\,m_{n}(x,y,t)$ requires then that $\int\!dxdy\,\int\!d\sigma
\,\sigma^{n}\phi\overline{O}=0$ and hence that,%
\begin{equation}%
%TCIMACRO{\dsum \limits_{k=2}^{\infty}}%
%BeginExpansion
{\displaystyle\sum\limits_{k=2}^{\infty}}
%EndExpansion
\frac{d\mu_{k}}{dt}\int\!dxdy\,h_{nk}=\nu\int\!dxdy\,h_{n2}\left\vert
\nabla\chi\right\vert ^{2},\ \ \ n=2,3,....\label{mu_Dot}%
\end{equation}
From this infinite set of equations, linear in $\overset{}{d\mu_{2}%
}/dt,\overset{}{d\mu_{3}}/dt,....$, the $d\mu_{k}/dt$ can be solved, in
principle. We have then a \emph{viscous} model with an infinite number of
conservation laws. Such a viscous model becomes physically more relevant by
making it compatible with a quasi-stationary distribution $\widetilde{\phi
}_{S}(\sigma,\psi(x,y))$ as given by Equation (\ref{w-psi}) with associated
$\Omega(\psi)$ relation at time $T_{S},$ i.e., at time $T_{S}$\ we can
associate with $\widetilde{\phi}_{S}(\sigma,\psi(x,y))$ a distribution
function as in Equation (\ref{ansatz}) which is a solution of the
time-evolution Equations (\ref{defO}) and (\ref{o}), with suitable initial
conditions such that $\phi^{\ast}(\sigma,x,y,T_{S})=\widetilde{\phi}%
(\sigma,\chi(x,y,T_{S}),T_{S})=\widetilde{\phi}_{S}(\sigma,\psi(x,y))$ with
$\chi(x,y,T_{S})=-\beta\psi(x,y)$ and with $\mu(\sigma,T_{S})=\mu(\sigma).$
The question of reconstructing the distribution function $\widetilde{\phi}%
_{S}(\sigma,\psi(x,y))$ from the experimental $\omega-\psi$ relation will be
addressed in Section \ref{reconstructing}.

\label{h-theorem}In concluding we want to remark that, as it is shown in
Appendix A, an H-theorem holds for the extremal distributions $\phi^{\ast
}(\sigma,x,y,t)$ in the case of conserved masses $m(\sigma,t)$ treated in this
paper. More precisely, it is for these distributions with conserved masses
that one can prove that $dS_{0}(t)/dt\geq0.$ Such an H-theorem does \emph{not}
hold for the other measures of mixing $S_{r}$ with $r\neq0$ which are defined
and discussed in Appendix A. Therefore, this result fixes the $0$--degree of
mixing $S_{0}$ defined in (\ref{so})-(\ref{largeSo}) as \emph{the} appropiate
measure of mixing.

\section{Fast-mixing condition\label{mixingSectn}}

In this paper we are mainly concerned with dynamical models such that
$\partial m(\sigma,t)/\partial t\equiv0,$ as treated in the previous
Subsection. In such a case, the inequalities in (\ref{muchFaster}) will always
be satisfied\footnote{We exclude the exceptional situation $\partial
s_{0}/\partial t=0.$} and assuming that the probability distribution $\phi$ is
of the extremal form given in Equation (\ref{ansatz}), we can make use of
equation (\ref{dSo/dt}) in order to write inequality (\ref{fasterE}) of the
fast mixing condition as,
\begin{equation}
\frac{1}{A}\int\!dx\,dy\,\left\langle \left\vert \nabla\ln\phi\right\vert
^{2}\right\rangle \gg\frac{1}{E}\int\!dx\,dy\,\omega^{2}.\label{IneqMax}%
\end{equation}
Two observations are in place: the viscosity drops out from the final
expression and the r.h.s. is totally determined by the macroscopic vorticity
field $\omega(x,y,t).$

Although the energy-related fast-mixing condition (\ref{fasterE}) does not
play a role in obtaining the time-dependent distributions $\phi_{M}%
(\sigma,x,y,t)=\phi^{\ast}(\sigma,x,y,t),$ it is still an interesting issue to
determine to which extent this fast-mixing condition is actually satisfied or
not by the $\phi^{\ast}(\sigma,x,y,t)$ distributions or more general
$\phi(\sigma,x,y,t)$ satisfying equations (\ref{defO})-(\ref{sigmaO}). In
fact, for high Reynolds' number one expects that after a time-interval of fast
mixing one arrives at a well-mixed probability distribution with
$E(T_{S})\lesssim E(t_{o}).$ In order to investigate this in more detail we
will 1) derive a lower bound to $\left\langle \left\vert \nabla\ln
\phi\right\vert ^{2}\right\rangle $ and 2) we will investigate the extrema of
the l.h.s. of equation (\ref{IneqMax}) taking into account the constraints
(\ref{normalztn}) and (\ref{vortDef}). As it will be seen, the special
distributions defined in equation (\ref{ansatz}) play a mayor role in both questions.

\subsection{Lower bound on $\left\langle \left\vert \nabla\ln\phi\right\vert
^{2}\right\rangle $}

In order to find a lower bound for $\left\langle \left\vert \nabla\ln
\phi\right\vert ^{2}\right\rangle $ in terms of $\left\vert \nabla
\omega\right\vert ^{2}$ we begin by noting that $\nabla\omega=\left\langle
\sigma\nabla\ln\phi\right\rangle $ can also be written as $\nabla
\omega=\left\langle \left(  \sigma-\omega\right)  \nabla\ln\phi\right\rangle $
because $\left\langle \nabla\ln\phi\right\rangle =0.$ Applying then the
Cauchy-Schwartz inequality to $\left\vert \nabla\omega\right\vert
^{2}=\left\vert \left\langle \left(  \sigma-\omega\right)  \nabla\ln
\phi\right\rangle \right\vert ^{2}$ leads to the desired lower bound,
\begin{align}
\left\vert \nabla\omega\right\vert ^{2}  & \leq\left\langle \left(
\sigma-\omega\right)  ^{2}\right\rangle \left\langle \left\vert \nabla\ln
\phi\right\vert ^{2}\right\rangle =K_{2}\left\langle \left\vert \nabla\ln
\phi\right\vert ^{2}\right\rangle ,\label{lowerBound}\\
\mathrm{i.e.,}\ \ \ \ \frac{\left\vert \nabla\omega\right\vert ^{2}}{K_{2}}  &
\leq\left\langle \left\vert \nabla\ln\phi\right\vert ^{2}\right\rangle
,\nonumber
\end{align}
where $K_{2}$\ is the centered second moment $K_{2}(x,y,t):=\int
\!d\sigma\,\left(  \sigma-\omega(x,y,t)\right)  ^{2}\phi(\sigma,x,y,t).$

The lower bound on $\left\langle \left\vert \nabla\ln\phi\right\vert
^{2}\right\rangle $ that we have just found means that the fast-mixing
condition ($\ref{IneqMax}$) holds whenever%
\[
\frac{1}{A}\int\!dx\,dy\,\frac{\left\vert \nabla\omega\right\vert ^{2}}{K_{2}%
}\gg\frac{1}{E}\int\!dx\,dy\,\omega^{2},
\]
irrespectively of all the higher-order moments with $n>2.$ Recall that
$\int\!dx\,dy\,\left\vert \nabla\omega\right\vert ^{2}$ is directly related to
the dissipation of the macroscopic enstrophy $\Gamma_{2},$ confer equation
(\ref{decay}) with $n=1$.

It is rather straightforward to determine the family of vorticity
distributions, let us call them $\phi_{LB}(\sigma,x,y,t),$ that reach the
lower bound in (\ref{lowerBound}), i.e., that satisfy $\left\langle \left\vert
\nabla\ln\phi_{LB}\right\vert ^{2}\right\rangle =K_{2}^{-1}\left\vert
\nabla\omega\right\vert ^{2}.$ In Appendix \ref{appndC} it is proved that
\[
\phi_{LB}(\sigma,x,y,t)=\phi^{\ast}(\sigma,x,y,t),
\]
with $\phi^{\ast}(\sigma,x,y,t)$ as given by equation (\ref{ansatz}). In the
present context, the input for the determination of the potential functions
$\mu(\sigma,t)$ and $\chi(x,y,t)$ are the first and second $\sigma-$moments,%
\begin{align}
\omega(x,y,t)  & =\int\!d\sigma\,\sigma\phi_{LB}(\sigma,x,y,t),\label{input}\\
\mathrm{and}\ \ K_{2}(x,y,t)  & =\int\!d\sigma\,\left(  \sigma-\omega
(x,y,t)\right)  ^{2}\phi_{LB}(\sigma,x,y,t).\nonumber
\end{align}

\subsection{Extrema of $\int\!dx\,dy\,\left\langle \left\vert \nabla\ln
\phi\right\vert ^{2}\right\rangle $}

In Appendix \ref{appendD} , we investigate the extrema of $\int
\!dx\,dy\,\left\langle \left\vert \nabla\ln\phi\right\vert ^{2}\right\rangle $
taking into account the $(x,y)$-dependent constraints (\ref{vortDef}) and
(\ref{normalztn}). It turns out that in order to obtain sensible solutions it
is necessary to constrain also the distribution's second moment $\left\langle
\sigma^{2}\right\rangle .$ We find that \emph{all} the extremizer
distributions, call them $\phi_{ext}(\sigma,x,y,t),$ are local \emph{minima}
of $\int\!dx\,dy\,\left\langle \left\vert \nabla\ln\phi\right\vert
^{2}\right\rangle $ and have the form $\phi_{ext}(\sigma,x,y,t)=\phi^{\ast
}(\sigma,x,y,t),$ with $\phi^{\ast}(\sigma,x,y,t)$\ as in equation
(\ref{ansatz}). This is in total agreement with the lower bound $\left\langle
\left\vert \nabla\ln\phi_{LB}\right\vert ^{2}\right\rangle $ found in the
previous Subsection, confer Appendix \ref{appndC}. 
In fact, as one can check,%
\[
\left\langle \left\vert \nabla\ln\phi_{ext}\right\vert ^{2}\right\rangle
=\frac{\left\vert \nabla\omega\right\vert ^{2}}{K_{2}},
\]
i.e., these extremal distributions $\phi_{ext}$ reach the lower bound
(\ref{lowerBound}). In the present case, the potential functions $\mu
(\sigma,t)$ and $\chi(x,y,t)$ should be determined from the input functions
$\omega(x,y,t)$ and $K_{2}(x,y,t),$ just as in equations (\ref{input}).

Summing up the results of the two last Subsections: 1) \emph{all}
distributions $\phi(\sigma,x,y,t)$ with first moment $\omega(x,y,t)$ and
second moment $m_{2}(x,y,t)=\omega^{2}(x,y,t)+K_{2}(x,y,t),$ satisfy
$\left\langle \left\vert \nabla\ln\phi\right\vert ^{2}\right\rangle
\geq\left\langle \left\vert \nabla\ln\phi_{LB}\right\vert ^{2}\right\rangle
=K_{2}^{-1}\left\vert \nabla\omega\right\vert ^{2}$ , \newline2) the lowest
possible value of $\left\langle \left\vert \nabla\ln\phi\right\vert
^{2}\right\rangle $ is achieved for $\phi_{LB}(\sigma,x,y,t)=\phi_{ext}%
(\sigma,x,y,t)=\phi^{\ast}(\sigma,x,y,t),$ with $\phi^{\ast}(\sigma
,x,y,t)$\ as in equation (\ref{ansatz}) and\newline\ 3) if $\int
\!dx\,dy\,\left(  \left\vert \nabla\omega\right\vert ^{2}/K_{2}\right)
\gg(A/E)\int\!dx\,dy\,\omega^{2},$ then the fast-mixing condition
($\ref{IneqMax}$) holds implying that the energy changes in the time interval
$[t_{o},T_{S}]$ are small.

\section{Reconstructing $\mu(\sigma)$ from experimental
data\label{reconstructing}}

Suppose that in an experiment one is given an initial vorticity field with its
corresponding energy $E_{o}$ and that one finds, from a time $T_{S}$ onwards,
a quasi-stationary vorticity field%
\begin{align*}
\omega_{S}(x,y)  & =\Omega(\psi(x,y)),\\
\mathrm{with}\ \ \ \ \psi_{\min}  & \leq\psi(x,y)\leq\psi_{\max},
\end{align*}
with a monotonic $\Omega(\psi)$ and an energy $E_{S}\leq E_{o}.$ As we show
below this experimental data can be used in order to determine the $\mu
(\sigma)$ potential occuring in the quasi-stationary distribution
$\widetilde{\phi}_{S}(\sigma,\psi).$ Once the distribution $\widetilde{\phi
}_{S}(\sigma,\psi)$ has been reconstructed from the experimental data we can
then associate with it a time-dependent distribution function $\phi^{\ast
}(\sigma,x,y,t)$ as given by equation (\ref{ansatz}) which is a solution of
the time-evolution Equation (\ref{defO}) with suitable initial conditions and
such that at time $t=T_{S}$ one has$\ \phi^{\ast}(\sigma,x,y,T_{S}%
)=\widetilde{\phi}_{M}(\sigma,\chi(x,y,T_{S}),T_{S})=\widetilde{\phi}%
_{S}(\sigma,\psi(x,y))$ with $\chi=-\beta\psi$ and with $\mu(\sigma,T_{S}%
)=\mu(\sigma).$ Here $\beta$ can be defined such that $M_{2}(T_{S}),$ the
second microscopic-vorticity moment of the QSS at time $T_{S},$ is equal to
the enstrophy $\Gamma_{2}^{0}$ of the initial vorticity field $\omega
(x,y,t_{o}).$ More explicitly, the condition $M_{2}(T_{S})=M_{2}(t_{0}%
)=\Gamma_{2}^{0}$ leads, also in the viscous case, to
\begin{equation}
\beta=-\frac{\int_{A}\!dx\,dy\,\left(  d\Omega/d\psi\right)  }{\Gamma_{2}%
^{0}-\Gamma_{2}^{S}},\label{beta}%
\end{equation}
see further the discussion in Section \ref{conclusions}.

The time-dependence of $\mu(\sigma,t)$ is chosen as in Subsection
\ref{fourCharacterztn}\ref{time-dependent}, i.e., such that all masses
$m(\sigma,t)$ are constant in time.

In order to determine the $\mu(\sigma)$ potential from the experimental
$\Omega(\psi)$, we notice that any monotonic function, like $\widetilde
{\Omega}(\chi),$ can be expressed as%
\begin{align*}
\widetilde{\Omega}(\chi)  & =\frac{d\ }{d\chi}\ln Z(\chi),\\
\mathrm{where}\ \ Z(\chi)  & =\int_{-\infty}^{\infty}\!d\sigma\,\exp\left[
\mu(\sigma)+\chi\sigma\right]  ,
\end{align*}
for some appropriate function $\mu(\sigma).$ From the above formula and
demanding that $\chi=-\beta\psi,$ it follows that it is possible to express
$Z(\chi)$ in terms of $\Omega(\psi)$,%
\[
Z(\chi)=\left.  \exp\left[  -\beta\int_{\psi_{\min}}^{\psi}d\psi^{\prime
}\,\Omega(\psi^{\prime})\right]  \right\vert _{\psi=-\chi/\beta},
\]
where $\beta$ is determined from the experimental $\Omega(\psi)$ relation
using equation (\ref{beta}). In order to illustrate this procedure, we now
treat a number of cases in which we can obtain $\mu(\sigma)$ from $\Omega
(\psi)$\ either analytically or approximately.

\subsection{Linear $\omega-\psi$ relation\label{linearO}}

The case of a linear $\omega-\psi$ scatter plot, i.e.,%
\begin{align*}
\Omega_{l}(\psi)  & =\alpha_{1}\psi,\\
\mathrm{or}\ \ \widetilde{\Omega}_{l}(\chi)  & =-\alpha_{1}\chi/\beta,
\end{align*}
can be treated rather straightforwardly. In this case one gets,%
\[
Z_{l}(\chi)=\exp\left[  -\frac{\alpha_{1}}{2\beta}\left(  \chi^{2}-\chi_{\min
}^{2}\right)  \right]  ,
\]
with $\chi_{\min}^{2}:=-\beta\psi_{\min}.$ As one can check, taking $\mu
_{l}(\sigma)$ such that%
\[
\exp\mu_{l}(\sigma):=\frac{1}{\sqrt{2\pi}}\sqrt{\left\vert \frac{\beta}%
{\alpha_{1}}\right\vert }\exp\left[  \frac{\alpha_{1}}{2\beta}\chi_{\min}%
^{2}\right]  \exp\left[  \frac{\beta}{2\alpha_{1}}\sigma^{2}\right]  ,
\]
with $\alpha_{1}/\beta<0,$ leads to the desired $Z_{l}(\chi),$\
\[
\int_{-\infty}^{\infty}\!d\sigma\,\exp\left[  \mu_{l}(\sigma)+\chi
\sigma\right]  =\exp\left[  \frac{\alpha_{1}}{2\beta}\chi_{\min}^{2}\right]
\exp\left[  -\frac{\alpha_{1}}{2\beta}\chi^{2}\right]  =Z_{l}(\chi).
\]
Since
\begin{align*}
\alpha_{1}/\beta & <0,\ \ \ \mathrm{and}\\
E_{S}  & =\int_{A}\!dx\,dy\,\omega_{S}\psi_{S}=\frac{1}{\alpha_{1}}\int
_{A}\!dx\,dy\,\omega_{S}^{2}>0,
\end{align*}
it follows that $\alpha_{1}$ is positive and $\beta$ is negative. The
corresponding probability density is then%
\[
\phi_{S}(\sigma,x,y)=\frac{1}{\sqrt{2\pi}}\sqrt{\left\vert \frac{\beta}%
{\alpha_{1}}\right\vert }\exp\left[  \frac{\beta}{2\alpha_{1}}\left(
\sigma-\alpha_{1}\psi(x,y)\right)  ^{2}\right]  ,
\]
i.e., a Gaussian centered on $\alpha_{1}\psi$ with a width $\sqrt{\alpha
_{1}/\left\vert \beta\right\vert }$ which has the form of (\ref{w-psi}) with
$\mu(\sigma)=\mu_{2}\sigma^{2}$ and $\mu_{2}=\beta/\left(  2\alpha_{1}\right)
.$ In the present case the expression (\ref{beta}) for $\beta$ reads,%
\begin{equation}
\beta=-\frac{\alpha_{1}A}{\Gamma_{2}(0)-\Gamma_{2}(T_{S})},\label{linBeta}%
\end{equation}
where $\Gamma_{2}(t)$ is the enstrophy $\int dxdy\,\omega^{2}(x,y,t).$

\subsection{General nonlinear $\omega-\psi$ relation}

In the more general nonlinear case we first notice that%
\[
\int_{-\infty}^{\infty}\!d\sigma\,\frac{d\mu(\sigma)}{d\sigma}\exp\left[
\mu(\sigma)+\chi\sigma\right]  =-\chi Z(\chi)+\left.  \exp\left[  \mu
(\sigma)+\chi\sigma\right]  \right\vert _{-\infty}^{+\infty}.
\]
The boundary terms vanish, so that, using the notation of equation
(\ref{averages}), one has
\begin{equation}
\left\langle \frac{d\mu(\sigma)}{d\sigma}\right\rangle =-\chi
(x,y).\label{Zparts}%
\end{equation}
Introducing into this equality the Taylor expansion of $\mu(\sigma),$%
\[
\mu(\sigma)=\sum_{k=2}\mu_{k}\sigma^{k},\ \ \ \ \mathrm{i.e.,}\ \ \ \frac
{d\mu(\sigma)}{d\sigma}=\sum_{k=2}k\mu_{k}\sigma^{k-1},
\]
where the coefficients $\mu_{k}$ are independent of $\chi$, one gets,%
\begin{equation}
\sum_{k=2}k\mu_{k}m_{k-1}(\chi)=-\chi.\label{mus}%
\end{equation}
Using the recursion operator $\mathcal{L}$ defined by (\ref{opertL}), this
equation can be rewritten as,%
\begin{equation}
\left[  \sum_{k=2}k\mu_{k}\mathcal{L}^{k-2}\right]  \circ\widetilde{\Omega
}=-\chi,\label{omegaN}%
\end{equation}
where $\mathcal{L}^{n+1}:=\mathcal{L}^{n}\circ\mathcal{L}$. Equation
(\ref{omegaN}) is a nonlinear differential equation in $\widetilde{\Omega
}(\chi) $ but it can be reduced to a \emph{linear} equation in the partition
function $Z(\chi).$ This simplification is achieved by multiplying both sides
of equation (\ref{omegaN}) by the partition function $Z(\chi),$ and noticing
that%
\begin{align}
\mathrm{for}\ \ n  & \geq1,\ \ Z(\chi)\mathcal{L}^{n}\circ\widetilde{\Omega
}=\frac{d^{n}\left(  Z(\chi)\widetilde{\Omega}\right)  }{d\chi^{n}}%
,\label{Ln}\\
\mathrm{and\ \ for\ \ }n  & =0,\mathrm{\ }\ \ Z(\chi)\widetilde{\Omega}%
=\frac{dZ}{d\chi}.\nonumber
\end{align}
One realizes then that (\ref{omegaN}) reduces to the following \emph{linear}
differential equation,
\begin{equation}
\sum_{k=2}k\mu_{k}\frac{d^{k-1}Z}{d\chi^{k-1}}=-\chi Z,\label{linearZ}%
\end{equation}
that the partition function $Z(\chi)$ must satisfy. In Subsection
\ref{finiteNRconservatn}\ref{2nd&4thMomntConservtn} and in Appendix
\ref{appendixE} we consider some cases in which Equation (\ref{linearZ}) can
be reduced to a finite-order differential equation.

\subsection{Slightly nonlinear $\omega-\psi$ relation}

It is often experimentally found that the $\omega-\psi$ plots satisfy
$\Omega(-\psi)\simeq-\Omega(\psi).$ The odd character of $\Omega(\psi)$
implies that $\mu(-\sigma)=\mu(\sigma),$ i.e.,%
\[
\mu(\sigma)=\mu_{2}\sigma^{2}+\mu_{4}\sigma^{4}+\mu_{6}\sigma^{6}+\cdots.
\]
Moreover, in many cases these plots are nearly linear. More specifically, in
the Taylor expansion for $\mu(\sigma)$ and in%
\begin{equation}
\widetilde{\Omega}(\chi)=f_{1}\chi+f_{3}\chi^{3}+f_{5}\chi^{5}%
+....,\ \ \mathrm{for}\ \ \left\vert \chi\right\vert \leq\chi_{\max
}\label{O_expand}%
\end{equation}
one has that in a substantial interval around $\psi=0$ or $\chi=0$,%
\begin{equation}
Z(\chi)=C\exp\left[  1+\frac{1}{2}f_{1}\chi^{2}+\frac{1}{4}f_{3}\chi^{4}%
+\frac{1}{6}f_{5}\chi^{6}+....\right]  ,\label{Z_chi}%
\end{equation}
with%
\begin{align*}
\left\vert f_{n+2}\right\vert \chi^{2}  & <\left\vert f_{n}\right\vert
\ \ \ \mathrm{for}\ \ \ n=1,3,5,....,\\
\mathrm{and}\ \ \ \ \mu_{n+2}  & <\mu_{2}\mu_{n}\ \ \ \mathrm{for}%
\ \ \ n=2,4,6,....
\end{align*}
Inserting these powers expansions of $Z(\chi)$ and $\mu(\sigma)$ into
(\ref{linearZ}) allows us to express the $\left\{  \mu_{n}\right\}  $ in terms
of the $\left\{  f_{n}\right\}  ,$ i.e., it allows us to determine the
probability density $\exp\mu(\sigma)$ from the experimentally known
scatter-plot $\widetilde{\Omega}(\chi).$ For example, retaining terms up to
$f_{5}$ and $f_{3}^{2}$\ in the Taylor expansion of \ Equation (\ref{linearZ})
one gets that%
\begin{align*}
\mu_{2}  & =-\frac{1}{2}f_{1}^{-1}-\frac{3}{2}f_{3}f_{1}^{-3}+\frac{15}%
{2}f_{5}f_{1}^{-4}-12f_{3}^{2}f_{1}^{-5},\\
\mu_{4}  & =\frac{1}{4}f_{3}f_{1}^{-4}-\frac{5}{2}f_{5}f_{1}^{-5}+\frac{21}%
{4}f_{3}^{2}f_{1}^{-6},\\
\mu_{6}  & =\frac{1}{6}f_{5}f_{1}^{-6}-\frac{1}{2}f_{3}^{2}f_{1}^{-7},\\
\mu_{8}  & =\mu_{10}=......=0.
\end{align*}

\section{Conservation of a finite number of total
moments\label{finiteNRconservatn}}

As we have seen in Subsection \ref{fourCharacterztn}\ref{time-dependent} the
masses $m(\sigma,t)$ are conserved if the time dependence of $\mu(\sigma,t)$
satisfies equation (\ref{dmu/dtBig}). For the sake of simplicity one can
demand less, for example a finite number of total moments $M_{n}$ for $2\leq
n\leq N$ will be conserved by requiring that $\mu(\sigma,t)=\sum_{i=2}^{N}%
\mu_{i}(t)\sigma^{i}+\sum_{i=N+1}^{\infty}\mu_{i}\sigma^{i}$ where the
$\mu_{i}$ with $i\geq N+1$\ are time independent and $d\mu_{2}/dt,\ldots
,\ d\mu_{N}/dt$ satisfy equation (\ref{mu_Dot}). In this case the derivation
of an H-theorem as given in Appendix \ref{appendA} does not go through even if
one assumes that $\mu_{i}=0,$ for $i\geq N+1,$ and consequently that
$\int\!d\sigma dxdy\,\phi\overline{O}\ln\phi A=0$ holds, confer (\ref{dSo/dt}%
). This is so because the $m(\sigma,t)$ are not conserved and instead of
equation (\ref{dSo/dt}) one has that%
\[
\frac{dS_{0}(t)}{dt}=\frac{\nu}{A}\int\!d\sigma\int dx\,dy\,\frac{1}{\phi
}\left\vert \nabla\phi\right\vert ^{2}+\frac{1}{A}\int\!d\sigma\frac{\partial
m(\sigma,t)}{\partial t}\left(  \ln m(\sigma)+1\right)  .
\]
However, considering that fast mixing takes place at times $t<T_{S}$ for which
the changes in masses can be assumed to be small, we still may use equation
($\ref{IneqMax}$) as the fast-mixing condition. In the present case inequality
(\ref{muchFaster}) is not satisfied automatically but only in the following
weaker version%
\[
\left\vert \frac{\partial\ }{\partial t}\int\!d\sigma\,\sigma^{n}s_{0}%
(\sigma,t)\right\vert \gg\frac{1}{A}\left\vert \frac{dM_{n}(t)}{dt}\right\vert
,\ \ \mathrm{for}\ \ n\geq N+1.
\]

\subsection{Gaussian distributions conserving the second
moment\label{gaussionTime}}

It is instructive to consider the $N=2$ case, i.e.,
\[
\frac{\partial\mu(\sigma,t)}{\partial t}=\frac{d\mu_{2}(t)}{dt}\sigma^{2}.
\]
Then from (\ref{mu_Dot}) one gets that%
\begin{equation}
\frac{d\mu_{2}(t)}{dt}=\nu\frac{\int\!dxdy\,\left[  K_{2}^{-1}K_{3}^{2}%
+K_{2}^{2}-K_{4}\right]  \left\vert \nabla\chi\right\vert ^{2}}{\int
\!dxdy\,\left[  K_{2}^{-1}K_{3}^{2}+K_{2}^{2}-K_{4}\right]  }.\label{dmu/dt}%
\end{equation}
If further $\mu_{k}=0$ for $k>2$ and $\mu_{2}<0,$ we have then that
$\left\langle \overline{O}\ln\phi\right\rangle =0$ and the corresponding
solution is a Gaussian distribution,
\[
\phi_{G}(\sigma,x,y,t)=\sqrt{\frac{\left\vert \mu_{2}(t)\right\vert }{\pi}%
}\exp\left[  \mu_{2}(t)\left(  \sigma+\frac{\chi(x,y,t)}{2\mu_{2}(t)}\right)
^{2}\right]  .
\]
In this case we have a linear $\omega-\psi$ relation and since $\left\langle
\sigma\right\rangle =\omega(x,y,t),$ it follows that $\chi(x,y,t)=-2\mu
_{2}(t)\omega(x,y,t),$ and $K_{2}(x,y,t)=-\left[  2\mu_{2}(t)\right]  ^{-1}%
.$The centered moments $K_{n}(x,y,t)$ of these Gaussian distributions are
independent of $(x,y)$, therefore the differential equation (\ref{dmu/dt})
satisfied by $\mu_{2}(t)$ reduces to,%
\begin{align}
\frac{d\mu_{2}}{dt}  & =\frac{\nu}{A}\int\!dxdy\,\left\vert \nabla
\chi\right\vert ^{2}\label{Gaussians}\\
& =\frac{4\nu}{A}\mu_{2}^{2}(t)\int\!dxdy\,\left\vert \nabla\omega\right\vert
^{2}.\nonumber
\end{align}
The solution is,%
\begin{align}
\mu_{2}(t)  & =\frac{\mu_{2}(0)}{\left[  1-2A^{-1}\mu_{2}(0)\Delta(t)\right]
},\label{mu(t)}\\
\mathrm{where}\ \ \Delta(t)  & :=2\nu\int_{0}^{t}\!ds~\int\!dxdy\,\left\vert
\nabla\omega\right\vert ^{2}\geq0.\nonumber
\end{align}
From equation (\ref{decay}) with $n=1,$ one sees that $\Delta\left(  t\right)
$ is the accumulated change in macroscopic enstrophy $\Gamma_{2}(t),$ i.e.,
$\Delta(t)=\left(  \Gamma_{2}(0)-\Gamma_{2}(t)\right)  .$ Since $\Delta\left(
t\right)  \geq0$ grows with time and $\mu_{2}(0)<0,$ it follows
that$\ \left\vert \mu_{2}(t)\right\vert $ decreases, i.e., the width
$1/\sqrt{\left\vert \mu_{2}(t)\right\vert }$ gets larger as time
increases.\newline If the initial distribution is sufficiently narrow so that
$\left\vert \mu_{2}(0)\right\vert \gg A\left[  2\Delta(\infty)\right]  ^{-1}$
then equation (\ref{mu(t)}) tells us that there must exist a time $T^{\prime}$
such that for all later times $t>T^{\prime}$
\[
\mathrm{for}\ ~t>T^{\prime},\ \ \ \mu_{2}(t)\simeq-A\left[  2\Delta(t)\right]
^{-1},
\]
i.e., the inverse width $\mu_{2}(t)$ becomes independent of its initial value
$\mu_{2}(0).$ Often $\Delta(\infty)\simeq\Gamma_{2}(0)$ and the width will
become independent of its initial value if $\left\vert \mu_{2}(0)\right\vert
\gg A\left[  2\Gamma_{2}(0)\right]  ^{-1},$ which depends only on the initial
conditions. In particular, for the $\delta-$type initial conditions as in
(\ref{Go}) one has that%
\[
\lim_{t\downarrow0}\mu_{2}(t)=-\infty,
\]
and Equation (\ref{mu(t)}) becomes%
\begin{equation}
\mu_{2}(t)=-\frac{A}{2\left(  \Gamma_{2}(0)-\Gamma_{2}(t)\right)
}.\label{mu2}%
\end{equation}
The condition for fast mixing in the present case reads,%
\[
\frac{2\left\vert \mu_{2}(0)\right\vert }{\left[  A-2\mu_{2}(0)\Delta
(t)\right]  }\int\!dxdy\,\left\vert \nabla\omega\right\vert ^{2}\gg\frac{1}%
{E}\int\!dxdy\,\omega^{2}.
\]
If the initial distribution is sufficiently narrow, i.e., $\left\vert \mu
_{2}(0)\right\vert \gg\left[  \Delta(\infty)\right]  ^{-1}$ then the condition
for fast mixing at $t>T^{\prime}$ becomes,%
\[
\frac{2\nu}{A}\int\!dxdy\,\left\vert \nabla\omega\right\vert ^{2}\gg
\frac{\Delta(t)}{E(t)}\int\!dxdy\,\omega^{2}.
\]

If the initial flow is unstable then $\left\vert \nabla\omega\right\vert ^{2}$
can become very large; in fact, these gradients will become larger the smaller
the viscosity $\nu.$ Accordingly one conjuctures that even in the limit of
vanishingly small viscosity one may have that%
\[
\lim_{\nu\rightarrow0}\frac{2\nu}{A}\int\!dxdy\,\left\vert \nabla
\omega\right\vert ^{2}>0,
\]
see the analysis presented in \cite{eyink}.

The corresponding dynamical picture would be as follows: due to chaotic mixing
$\Delta(t)\,$grows very quickly up to a time $T_{S}$ when $\omega(x,y,T_{S})$
becomes, to a good approximation, a function of the corresponding
stream-function, i.e., $\omega(x,y,T_{S})\simeq\Omega(\psi(x,y,T_{S})).$ At
this moment one has a stationary solution of the inviscid Euler equations, the
(weak) time dependence is due to viscous effects. If the $\omega-\psi$
relation is linear then the quasi-stationary state $\phi_{S}\left(
\sigma,x,y\right)  $ is in total agreement with the Gaussian solutions
$\phi_{G}\left(  \sigma,x,y,T_{S}\right)  $ provided that the initial
distribution is sharp enough, i.e., that $\left\vert \mu_{2}(0)\right\vert \gg
A\left[  2\Delta(\infty)\right]  ^{-1}$ and that $T_{S}>T^{\prime}.$\ In fact,
from the linear $\omega-\psi$ relation $\omega=\alpha_{1}\psi$ of Section
\ref{reconstructing}A with $\beta$ satisfying (\ref{beta}) it follows that,%
\[
\alpha_{1}\psi=-\beta^{-1}\alpha_{1}\chi=-A^{-1}\left(  \Gamma_{2}%
(0)-\Gamma_{2}(T_{S})\right)  \chi=\left(  2\mu_{2}(T_{S})\right)  ^{-1}\chi.
\]

The Gaussian distributions that we have discussed above conserve only the
distribtuion's second moment. Suppose that at a certain instant $T$ we have a
Gaussian distribution, i.e., that $\mu(\sigma,T)=\mu_{2}\sigma^{2},$ can this
distribution conserve all moments and stay Gaussian at later times $t>T?$
Equation (\ref{dmu/dtBig}) tells us that for this to be true the following
identity should hold,%
\begin{align*}
\int\left[  \frac{d\mu_{2}}{dt}\left[  \sigma^{2}-\left\langle \sigma
^{2}\right\rangle -2\omega(\sigma-\omega)\right]  \right]  \phi(\sigma
,x,y,t)\,dxdy  & =\\
-\nu\int\left[  K_{2}-\left(  \sigma-\omega\right)  ^{2}\right]  \left\vert
\nabla\chi\right\vert ^{2}\phi(\sigma,x,y,t)\,dxdy  & =,
\end{align*}
or, after some manipulations, that,%
\begin{align*}
\frac{d\mu_{2}}{dt}\int\left[  \left(  \sigma-\omega\right)  ^{2}%
-K_{2}\right]  \phi(\sigma,x,y,t)\,dxdy  & =\\
\nu\int\left[  \left(  \sigma-\omega\right)  ^{2}-K_{2}\right]  \left\vert
\nabla\chi\right\vert ^{2}\phi(\sigma,x,y,t)\,dxdy  & =
\end{align*}
Therefore $\left\vert \nabla\chi\right\vert ^{2}$ must be independent of
$\left(  x,y\right)  $ and so must be $\left\vert \nabla\omega\right\vert ^{2}
$ since $\chi=2\mu_{2}\omega.$ This corresponds to the simplest dynamical
model, equation (\ref{simple}) with $\overline{O}\equiv0.$ Therefore, a
Gaussian distribution can conserve the masses $m(\sigma,t)$ only in the very
special case of a space independent $\left\vert \nabla\omega\right\vert ^{2}.$
Moreover, the last equation implies that,%
\begin{align*}
\frac{d\mu_{2}}{dt}  & =\nu\left\vert \nabla\chi\right\vert ^{2}\\
& =4\nu\mu_{2}^{2}(t)\left\vert \nabla\omega\right\vert ^{2},
\end{align*}
in agreement with our previous results, confer Equation (\ref{Gaussians}).

\subsection{More general distributions with conserved second
moment\label{oneScale}}

Here we study the family of nonlinear $\mu(\sigma,t)$ with only one
time-dependent $\sigma$-scale $q(t),$ i.e.,%
\[
\mu(\sigma,t)=\widetilde{\mu}(q(t)\sigma),
\]
and accordingly,%
\begin{align}
\frac{\partial\mu(\sigma,t)}{\partial t}  & =\widetilde{\mu}^{\prime}\frac
{dq}{dt}\sigma\label{scales}\\
& =\frac{\sigma}{q}\frac{\partial\mu}{\partial\sigma}\frac{dq}{dt},\nonumber
\end{align}
where $\widetilde{\mu}^{\prime}:=\left.  d\widetilde{\mu}(x)/dx\right\vert
_{x=q\sigma}$ and the time-dependence of $q(t)$ will be determined in the
following paragraphs such that the second global moment is conserved, i.e.,
$M_{2}(t)=\int\!dx\,dy\,\left\langle \sigma^{2}\right\rangle =M_{2}(0).$

The recursion operator $\mathcal{L}$ defined in (\ref{opertL}) allows us to
write%
\[
\left\langle \sigma\frac{d\mu}{d\sigma}\right\rangle =\mathcal{L}\left\langle
\frac{d\mu}{d\sigma}\right\rangle =-\mathcal{L}{\normalsize \chi,}%
\]
confer Equation (\ref{Zparts}). After some algebra, it follows from
(\ref{scales}) that,%
\begin{align*}
\nu\left\langle \Gamma\right\rangle  & =-\left(  1+\chi\widetilde{\Omega}%
(\chi)\right)  \frac{1}{q}\frac{dq}{dt},\\
\nu\left\langle \left(  \sigma-\omega\right)  \Gamma\right\rangle  & =-\left(
\widetilde{\Omega}+\chi\frac{d\widetilde{\Omega}}{d\chi}\right)  \frac{1}%
{q}\frac{dq}{dt},\\
\mathrm{and}\ \ \ \nu\left\langle \sigma^{2}\Gamma\right\rangle  & =-\frac
{1}{q}\frac{dq}{dt}\left[  \chi\frac{d^{2}\widetilde{\Omega}}{d\chi^{2}%
}+3\left(  1+\chi\Omega\right)  \frac{d\widetilde{\Omega}}{d\chi}+3\Omega
^{2}+\chi\Omega^{3}\right]  .
\end{align*}
Multiplying equation (\ref{o}) by $\sigma^{2},$ taking the average over the
vorticity distribution and using the expressions we just derived one arrives
at,%
\[
\nu\left\langle \sigma^{2}\overline{O}\right\rangle =\nu\left\vert \nabla
\chi\right\vert ^{2}\left[  \frac{K_{3}^{2}}{K_{2}}+K_{2}^{2}-K_{4}\right]
-\frac{1}{q}\frac{dq}{dt}\left[  2K_{2}-\Omega\frac{K_{3}}{K_{2}}\right]  .
\]
The conservation of the second global moment $M_{2}=\int\!dx\,dy\,\left\langle
\sigma^{2}\right\rangle $ requires that $\int\!dx\,dy\,\left\langle \sigma
^{2}\overline{O}\right\rangle =0,$ it follows then that $M_{2}(t)$ is
conserved if,%
\[
\frac{1}{q}\frac{dq}{dt}=\nu\frac{\int\!dx\,dy\,\left\vert \nabla
\chi\right\vert ^{2}K_{2}^{-1}\left[  K_{3}^{2}+K_{2}^{3}-K_{2}K_{4}\right]
}{\int\!dx\,dy\,K_{2}^{-1}\left[  2K_{2}^{2}-\Omega K_{3}\right]  }.
\]
In most cases one will have that $\left[  K_{3}^{2}+K_{2}^{3}-K_{2}%
K_{4}\right]  <0$ while $\left[  2K_{2}^{2}-\Omega K_{3}\right]  >0,$
therefore, the inverse width $q^{2}$\ will decrease in time, i.e., the width
of the distribution will grow in time. If the distribution is Gaussian then
$K_{3}=0,$ $K_{4}=3K_{2}^{2}$ , $K_{2}=-\left[  2\mu_{2}(t)\right]  ^{-1}$ and
the last expression becomes,%
\begin{align*}
\frac{1}{q}\frac{dq}{dt}  & =-\nu\frac{\int\!dx\,dy\,\left\vert \nabla
\chi\right\vert ^{2}K_{2}^{2}}{\int\!dx\,dy\,K_{2}}\\
& =-\nu\frac{K_{2}}{A}\int\!dx\,dy\,\left\vert \nabla\chi\right\vert ^{2}.
\end{align*}
This is in agreement with the result given in equation (\ref{Gaussians}).

If $d\widetilde{\mu}(x)/dx$\ is a nonlinear function of $x$ then it does not
follow that $\int\!d\sigma dxdy\,\phi\overline{O}\ln\left(  \phi A\right)  =0$
and there is no H-theorem, confer Equation (\ref{A}). Assume now that at time
$T_{S}$ there is a nonlinear $\omega$--$\psi$ relation which can be associated
with a distribution of type (\ref{ansatz}) with a $\mu(\sigma,t)$ as in
equation (\ref{scales}). We can choose the value of $\beta$ according to
(\ref{beta}) so that $\delta_{2}(0)=0.$ In the context of our non-Gaussian
distributions with conserved second moment this means that the distribution
$\phi(\sigma,x,y,T_{S})=\widetilde{\phi}_{S}(\sigma,\psi(x,y))$ can be
obtained from a $\delta$-function like initial distribution.

\subsection{Non-Gaussian distributions with conserved second and fourth
moments\label{2nd&4thMomntConservtn}}

Let us now consider the special case when in the Taylor expansion of
$\mu(\sigma)$ only $\mu_{2}$ and $\mu_{4}$ do not vanish. Then the
differential equation (\ref{linearZ}) satisfied by the corresponding $Z\left(
\chi\right)  $ reduces to,
\begin{equation}
2\mu_{2}\frac{dZ}{d\chi}+4\mu_{4}\frac{d^{3}Z}{d\chi^{3}}=-\chi
Z\ \ \ \mathrm{with}\ \ \mu_{4}<0,\label{cubic}%
\end{equation}
From this equation one determines the asymptotic behaviour of $\widetilde
{\Omega}(\chi)=d\ln Z/d\chi$ as%
\begin{equation}
\widetilde{\Omega}(\chi)\underset{\chi\rightarrow\infty}{\rightarrow}\frac
{4}{3}\left(  \frac{\chi}{4\left\vert \mu_{4}\right\vert }\right)
^{1/3}+\frac{\mu_{2}}{6\left\vert \mu_{4}\right\vert }\left(  \frac{\chi
}{4\left\vert \mu_{4}\right\vert }\right)  ^{-1/3}+O\left(  \frac{4\left\vert
\mu_{4}\right\vert }{\chi}\right)  .\label{xi_limit}%
\end{equation}
We see that in this case $\lim\widetilde{\Omega}(\chi)\rightarrow\pm\infty$
for $\chi\rightarrow\pm\infty.$ In fact, from Equation (\ref{linearZ}) one can
prove that $\widetilde{\Omega}(\chi)$\ remains finite in the limit $\left\vert
\chi\right\vert \rightarrow\infty$ iff infinitely many $\mu_{k}$\ do not
vanish. Using equation (\ref{cubic}) one can express the coefficients $\mu
_{2}$ and $\mu_{4}$ in terms of the experimentally determined coefficients of
$\widetilde{\Omega}(\chi)=f_{1}\chi+f_{3}\chi^{3}+f_{5}\chi^{5}+.....$
Inserting this expression in equation (\ref{cubic}), expanding in powers of
$\chi$ and equating the coefficients of $\chi^{2}$ and $\chi^{4}$\ on both
sides of the resulting equality leads to%
\begin{align*}
\mu_{2}  & =-\frac{2f_{1}^{3}+24f_{1}f_{3}+40f_{5}}{4f_{1}^{4}+36f_{1}%
^{2}f_{3}+80f_{1}f_{5}-24f_{3}^{2}},\\
\mu_{4}  & =\frac{f_{3}}{4f_{1}^{4}+36f_{1}^{2}f_{3}+80f_{1}f_{5}-24f_{3}^{2}%
}.
\end{align*}
\newline In the special case $\mu_{2}=0$ the solution of equation
(\ref{cubic}) with boundary conditions $Z(0)=1,\,Z^{\prime}(0)=0$ and
$Z^{\prime\prime}(0)=0$ is the Meijer G-function or hypergeometric function
$_{0}F_{2}([\frac{1}{2}\,,\,\frac{3}{4}],\chi^{4}/256\left\vert \mu
_{4}\right\vert ).$ The corresponding $\widetilde{\Omega}(\chi)$ is plotted in
Fig. 1. As expected for this special case with $\mu_{2}=0$, around $\chi=0$
the function $\widetilde{\Omega}(\chi)$ is not linear but cubic, i.e.,
$f_{1}=0$.
%TCIMACRO{\FRAME{ftbpFU}{4.7556in}{7.3016in}{0pt}{\Qcb{Plot of $\left\vert
%\mu_{4}\right\vert ^{1/4}\widetilde{\Omega}$ as a function of $\chi/\left\vert
%\mu_{4}\right\vert ^{1/4}$ for the case $\mu(\sigma)=\mu_{4}\sigma^{4}.$}%
%}{\Qlb{fig1}}{mu4.eps}{\special{ language "Scientific Word";  type "GRAPHIC";
%width 4.7556in;  height 7.3016in;  depth 0pt;  original-width 6.4913in;
%original-height 9.9877in;  cropleft "0";  croptop "1";  cropright "1";
%filename '../../Program Files/Maple 8/MyMaple/mu4.eps';file-properties "XNPEU";}%
%}}%
%BeginExpansion
\begin{figure}
[ptb]
\begin{center}
\includegraphics[
height=7.3016in,
width=4.7556in
]%
{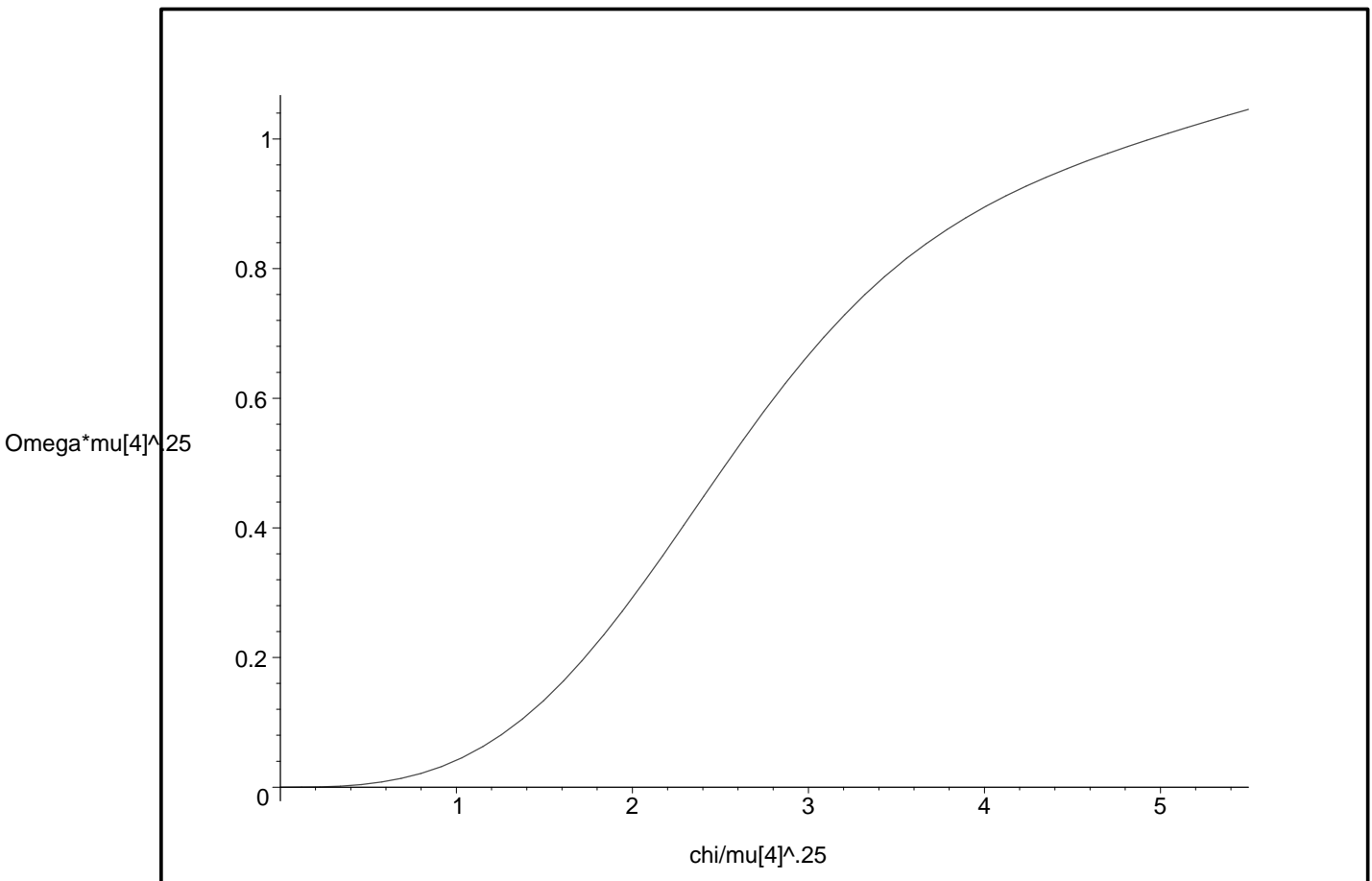}%
\caption{Plot of $\left\vert \mu_{4}\right\vert ^{1/4}\widetilde{\Omega}$ as a
function of $\chi/\left\vert \mu_{4}\right\vert ^{1/4}$ for the case
$\mu(\sigma)=\mu_{4}\sigma^{4}.$}%
\label{fig1}%
\end{center}
\end{figure}
%EndExpansion

The coefficients $\mu_{2}$ and $\mu_{4}$ in the case of equation (\ref{cubic})
can be time-dependent; their time-dependence can be fixed, e.g., by requiring
that the second and fourth global moments $M_{2}$ and $M_{4}$ be conserved.
Such requirement, i.e., equations (\ref{mu_Dot}) with $n=2,4$ and $\mu_{n}=0$
for $n\neq2,4$ leads to%
\begin{align*}
\frac{d\mu_{4}}{dt}  & =\nu\left[  \frac{H_{22}\int dxdy\,\left\vert
\nabla\chi\right\vert ^{2}h_{42}-H_{42}\int dxdy\,\left\vert \nabla
\chi\right\vert ^{2}h_{22}}{h_{22}h_{44}-h_{24}h_{42}}\right]  ,\\
\frac{d\mu_{2}}{dt}  & =\nu\frac{\int dxdy\,\left\vert \nabla\chi\right\vert
^{2}h_{22}}{H_{22}}-\frac{d\mu_{4}}{dt}\frac{H_{24}}{H_{22}},\\
\mathrm{where}\ \ \ \ H_{nm}  & :=\int dxdy\,h_{nm}%
\end{align*}
and the relation $m_{n}=\sum\limits_{m=0}^{n}\left(
\begin{array}
[c]{c}%
n\\
m
\end{array}
\right)  \omega^{m}K_{n-m}$ so that the time-derivatives can be expressed as
complicated functions of $K_{2},K_{3},....,K_{8}.$ It is instructive to
consider a nearly Gaussian situation in which the $K_{n}$ are replaced by
their Gaussian values $K_{2m-1}=0$ and $K_{2m}=(2m-1)!!K_{2}^{m}.$ Then%
\begin{align*}
h_{22}  & =2K_{2}^{2},\ \ h_{42}=12K_{2}^{2}(K_{2}+\omega^{2}),\\
h_{44}  & =24\left(  K_{2}^{2}+10K_{2}\omega^{2}+3\omega^{4}\right)  ,\\
\frac{d\mu_{4}}{dt}  & =\frac{\nu}{2}\frac{\int dxdy\,\left\vert \nabla
\omega\right\vert ^{2}\left(  \omega^{2}-\Gamma_{2}A^{-1}\right)  K_{2}^{-2}%
}{AK_{2}^{2}+4K_{2}\Gamma_{2}+3(\Gamma_{4}-\Gamma_{2}^{2})},\\
\frac{d\mu_{2}}{dt}  & =\frac{\nu}{AK_{2}^{2}}\int dxdy\,\left\vert
\nabla\omega\right\vert ^{2}-6\left(  K_{2}+\Gamma_{2}A^{-1}\right)
\frac{d\mu_{4}}{dt}.
\end{align*}
The expression for $d\mu_{4}/dt$ is governed by the correlation between
$\left\vert \nabla\omega\right\vert ^{2}$ and $\omega^{2}.$ Close to the
maxima of $\omega$ the gradient will be very small, analogously the gradient
maybe large in the areas where $\omega$ is not close to an extremum.
Therefore, it is reasonable to assume that this correlation will be negative,
so that $d\mu_{4}/dt\,<0$ while $d\mu_{2}/dt>0.$ This means that, in the
context of the models with conserved second and fourth moments, a slightly
non-Gaussian initial distribution will evolve into a more strongly
non-Gaussian distribution.

\section{Discussion and conclusions\label{conclusions}}

In the MRS approach the conservation laws of 2D inviscid flows play a central
role. However, the unavoidable viscosity makes these conservation laws
invalid; in fact, at high Reynolds' numbers it is only the energy that is
approximately conserved\cite{matthaeSelect}. On the other hand, as we have
seen, the conservation of the microscopic-vorticity masses $m(\sigma,t) $ is
compatible with a non-vanishing viscosity, it leads to spatial mixing of the
masses without destroying or creating them. This is the picture adopted in the
present paper.

By working with the family of maximally mixed states studied in Sections
\ref{Mixing}-\ref{mixingSectn} we reached a mathematical formulation that, as
far as the QSS is concerned, shows many parallels with the MRS formulation. In
Section \ref{reconstructing}\ we addressed the problem of how to determine the
QSS distribution $\phi_{S}(\sigma,x,y)$ from an experimental $\omega-\psi$
relation and $\beta$ given by Equation (\ref{beta}). Identifying this
$\phi_{S}(\sigma,x,y)$ with the distribution $\phi(\sigma,x,y,T_{S})=\phi
_{M}(\sigma,x,y,T_{S})=\phi_{LB}(\sigma,x,y,T_{S})=\phi^{\ast}(\sigma
,x,y,T_{S})$ of Section \ref{fourCharacterztn} and using a time-dependent
$\mu(\sigma,t)$ satisfying Equations (\ref{dmu/dtBig})-(\ref{mu_Dot}) and such
that $\mu(\sigma,T_{S})=\mu(\sigma)$ we have a dynamical model that conserves
the masses, i.e. with $m(\sigma,t)=m(\sigma,t_{o}),$ and connects the
experimental $\omega-\psi$ relation found at time $T_{S}$ with an initial
condition $\phi(\sigma,x,y,t_{o}).$ An extra bonus that follows from this
methodology is that an H-theorem holds, moreover, it holds for one measure of
spatial mixing, namely for the $S_{0}$ degree of mixing, confer Appendix A.
When the masses $m(\sigma,t)$ are conserved, this mixing measure $S_{0}$
coincides with the MRS entropy.

In Subsection III B of an earlier paper\cite{Capel} we expressed the
quantities $\delta_{n}$
\[
\delta_{n}:=\int\!d\sigma\int\!dxdy\,\left[  \sigma^{n}-\omega_{S}%
^{n}(x,y)\right]  \phi_{S}(\sigma,x,y),
\]
in terms of spatial integrals of certain polynomials in $\Omega(\psi)$ and its
derivatives $d^{r}\Omega/d\psi^{r}.$ For example, for $n=2,$ one has that%
\[
\delta_{2}=-\frac{1}{\beta}\int_{A}\!dx\,dy\,\frac{d\Omega}{d\psi}.
\]
and the MRS assumption $0=\delta_{2}+\Gamma_{2}^{S}-\Gamma_{2}^{0}$, can be
used, as we did in Section \ref{reconstructing}, in order to determine $\beta$
from the experimentally accessible quantities $\Omega(\psi)$ and the initial
and final enstrophies $\Gamma_{2}^{0}$ and $\Gamma_{2}^{S}$ as in Equation
(\ref{beta}). From these $\delta_{n}$ one finds the corresponding initial
values%
\[
\delta_{n}^{o}:=\int\!d\sigma\int\!dxdy\,\left[  \sigma^{n}-\omega_{o}%
^{n}(x,y)\right]  \phi(\sigma,x,y,t_{o}),
\]
because the conservation of the moments $M_{n}$ implies that$\ \delta_{n}%
^{o}=\delta_{n}+\Gamma_{n}^{S}-\Gamma_{n}^{0},$ where $\Gamma_{n}^{S}%
:=\int\!dx\,dy\,\omega_{S}^{n}(x,y)$ and $\Gamma_{n}^{0}:=\int\!dx\,dy\,\omega
_{o}^{n}(x,y)$ are the macroscopic-vorticity moments in the QSS and in the
initial state, repectively. In the MRS approach the initial distribution is
assumed to be a $\delta$-function as in Equation (\ref{Go}), consequently, all
$\delta_{n}^{o}$ should vanish. This led us in \cite{Capel} to propose the
quantities,%
\begin{equation}
\frac{\delta_{n}}{\Gamma_{n}^{0}-\Gamma_{n}^{S}}%
,\ \ n=2,3,....\label{yardsticks}%
\end{equation}
as yardsticks in order to investigate the validity of the statistical
mechanical approach according to which all these quantities should equal 1.
Choosing $\beta$ as in (\ref{beta}) the yardstick relation $\delta_{2}/\left(
\Gamma_{2}^{0}-\Gamma_{2}^{S}\right)  =1$ is automatically satisfied while the
remaining yardsticks for $n>2$ are nontrivial checks for the validity of the
statistical mechanics approach. When all the yardsticks relations $\delta
_{n}/(\Gamma_{n}^{o}-\Gamma_{n})=1$ hold the quasi-stationary state predicted
by the MRS approach is in agreement with the experimental $\omega-\psi$
relation, moreover, it is also the solution of equation (\ref{defO}) at time
$T_{S}$ with conserved total moments $M_{n}$ and starting from the initial
condition $\phi(\sigma,x,y,t_{o})=\delta(\sigma-\omega_{0}(x,y)).$ When not
all the yardsticks relations are satisfied then the MRS approach gives an
incorrect but approximate prediction of the experimental $\Omega(\psi)$
relation. In such a case it still may be possible to obtain a rather sharp
initial distribution, which is determined by the experimental $\Omega(\psi)$
relation, such that the quasi-stationary state is the solution of equation
(\ref{defO}) at time $T_{S}$ and with conserved total moments $M_{n}%
(T_{S})=M_{n}(t_{o}).$

Hence it maybe concluded that our class of dynamical models with non-vanishing
viscosity and with conserved masses yields a self-consistent dynamical
description containing the MRS approach with $\delta-$type initial conditions
in the case that the experimental values of the yardsticks relations are 1 and
with different initial conditions in all other cases. In these other cases the
predictive power of the statistical mechanics approach, i.e., the prediction
of the correct QSS on the basis of a $\delta-$type initial condition, is lost
and the reconstruction of the correct initial distribution in the context of
the models with conserved masses is a rather difficult if not forbidding
problem. In such cases it may be more convenient to give up the conservation
of all masses, require that only the second moment $M_{2}$ be conserved and
use the Gaussian distributions of Subsection \ref{finiteNRconservatn}%
\ref{gaussionTime} if the $\omega-\psi$ relation is linear or the family of
$\mu(\sigma,t)$ with only one time-dependent scale $q(t)$ of Subsection
\ref{finiteNRconservatn}\ref{oneScale} in the case of a nonlinear $\omega
-\psi$ relation.

\newpage

\appendix

\section{Degrees of mixing\label{appendA}}

A mass of solute $m$ achieves the highest degree of mixing when it is
homogeneously distributed over the area $A,$ i.e., when its concentration is
constant and equal to $m/A.$ Let us introduce $\delta(\sigma,x,y,t)$ the
spatial distribution of the $\sigma$-species at time $t,$ i.e.,
\begin{align*}
\delta(\sigma,x,y,t) &  :=\frac{\phi(\sigma,x,y,t)}{m(\sigma,t)}\geq0,\\
\int\!dx\,dy\,\delta(\sigma,x,y,t) &  =1.
\end{align*}
In order to determine how well mixed is this $\sigma$-mass one has to
determine `how close' is the corresponding $\delta(\sigma,x,y,t)$ to the
homogeneous distribution $1/A.$ As it is known\cite{arndt}, given two spatial
densities, called them $\delta_{1}(x,y)$ and $\delta_{2}(x,y),$ there is a
one-parameter family of non-negative, convex functionals $d_{r}(\delta
_{1},\delta_{2})$ satisfying $d_{r}(\delta_{1},\delta_{2}%
)=0\longleftrightarrow\delta_{1}\overset{a.e.}{=}\delta_{2}$ and measuring a
sort of distance between them, namely
\[
d_{r}(\delta_{1},\delta_{2}):=\left[  r(1-r)\right]  ^{-1}\int\!dx\,dy\,\delta
_{1}(x,y)\left[  1-\left(  \delta_{2}(x,y)/\delta_{1}(x,y)\right)
^{r}\right]  \geq0,\;0<r<1,
\]
and, by imposing continuity in $r,$
\begin{align*}
d_{0}(\delta_{1},\delta_{2}) &  :=-\int\!dx\,dy\,\delta_{1}(x,y)\ln\left(
\delta_{2}(x,y)/\delta_{1}(x,y)\right)  ,\\
d_{1}(\delta_{1},\delta_{2}) &  :=-\int\!dx\,dy\,\delta_{2}(x,y)\ln\left(
\delta_{1}(x,y)/\delta_{2}(x,y)\right)  .
\end{align*}
For $r>1$ and for $r<0$, they may diverge. However, since in our case
$\delta_{2}(x,y)=1/A\neq0,$ values $r<0$\ would also be possible.\ The $d_{r}$
are not distances because, e.g., for $r\neq1/2,$ they are asymmetric in their
arguments, i.e., $d_{r}(\delta_{1},\delta_{2})\neq d_{r}(\delta_{2},\delta
_{1}).$

Accordingly, we can measure the mixing degree of the $\sigma$-mass by
$-d_{r}(\delta(\sigma,x,y,t),1/A).$ Its weighted contribution to the total
mixing degree will be denoted by $s_{r}(\sigma,t),$ i.e.,
\begin{align*}
s_{r}(\sigma,t) &  :=m(\sigma,t)\left[  Ar(r-1)\right]  ^{-1}\int
\!dx\,dy\,\delta(\sigma,x,y,t)\left[  1-\left(  1/A\delta(\sigma
,x,y,t)\right)  ^{r}\right]  \leq0,\;\mathrm{for}\;0<r<1,\\
s_{0}(\sigma,t) &  :=-m(\sigma,t)A^{-1}\int\!dx\,dy\,\delta(\sigma,x,y,t)\ln
A\delta(\sigma,x,y,t)\\
&  =-A^{-1}\int\!dx\,dy\,\phi(\sigma,x,y,t)\ln\left[  \frac{A\phi
(\sigma,x,y,t)}{m(\sigma,t)}\right]  \leq0,\\
s_{1}(\sigma,t) &  :=m(\sigma,t)A^{-1}\int\!dx\,dy\,\ln A\delta(\sigma
,x,y,t)\leq0,
\end{align*}
the corresponding total $r$-order degree of mixing is then,
\begin{align}
S_{r}(t) &  :=\left[  Ar(r-1)\right]  ^{-1}\int\!d\sigma\int\!dx\,dy\,\phi
(\sigma,x,y,t)\left[  1-\left(  m(\sigma,t)/A\phi(\sigma,x,y,t)\right)
^{r}\right]  \leq0,\;0<r<1,\label{degreeofmix}\\
S_{0}(t) &  :=-A^{-1}\int\!d\sigma\int\!dx\,dy\,\phi(\sigma,x,y,t)\ln\left[
A\phi(\sigma,x,y,t)/m(\sigma,t)\right]  \leq0.\nonumber
\end{align}

Under certain conditions, these degrees of mixing satisfy a kind of H-theorem.
In order to see this, let us compute the time derivative of $s_{r}(\sigma,t),$
with $0<r<1,$%
\[
\frac{\partial s_{r}(\sigma,t)}{\partial t}=\frac{1}{Ar\left(  r-1\right)
}\int\!dx\,dy\,\left\{  \left[  1+\frac{\left(  r-1\right)  m^{r}}{\left(
A\phi\right)  ^{r}}\right]  \frac{\partial\phi}{\partial t}-\frac{r}{A}\left(
\frac{m}{A\phi}\right)  ^{(r-1)}\frac{\partial m}{\partial t}\right\}  ,
\]
and consider the mass-consevation models with $\partial m/\partial t=0$\ so
that $\partial s_{r}(\sigma,t)/\partial t$ is totally determined by the first
term in the curly brackets and one has that,%
\begin{align}
\frac{\partial s_{r}(\sigma,t)}{\partial t} &  =\frac{1}{r\left(  r-1\right)
A}\int\!dx\,dy\,\left\{  \left[  1+\frac{\left(  r-1\right)  m^{r}(\sigma
)}{\left(  A\phi\right)  ^{r}}\right]  \left[  -\vec{v}\cdot\nabla\phi
+\nu\Delta\phi+\nu\overline{O}\phi\right]  \right\} \\
&  =\frac{\nu}{A}\int\!dx\,dy\,\frac{m^{r}(\sigma)}{\left(  A\phi\right)
^{r}}\left[  \frac{\left\vert \nabla\phi\right\vert ^{2}}{\phi}+\frac{1}%
{r}\overline{O}\phi\right]  \ \ \mathrm{with}\;0<r<1,
\end{align}
While for $r=0,$%
\begin{align}
\frac{\partial s_{0}(\sigma,t)}{\partial t}  & =\frac{\nu}{A}\int
\!dx\,dy\,\phi\left[  \left\vert \frac{\nabla\phi}{\phi}\right\vert
^{2}+\overline{O}\ln\frac{m(\sigma)}{A\phi}\right] \label{A}\\
& =\frac{\nu}{A}\int\!dx\,dy\,\phi\left[  \left\vert \frac{\nabla\phi}{\phi
}\right\vert ^{2}-\overline{O}\ln A\phi\right]  .\nonumber
\end{align}

In the simplest viscous model with $\overline{O}=0,$ the last formulas lead to
a H-theorem for each $\sigma$-value,%
\[
\frac{\partial s_{r}(\sigma,t)}{\partial t}=\frac{\nu}{A}\int\!dx\,dy\,\frac
{m^{r}(\sigma)}{\left(  A\phi\right)  ^{r}}\frac{\left\vert \nabla
\phi\right\vert ^{2}}{\phi}\geq0,\ \mathrm{with}\;0\leq r\leq1.
\]
In the more general viscous model with $\overline{O}\neq0$ and conservation of
the total moments, i.e., with $\int\!dx\,dy\,\left\langle \sigma^{n}%
\overline{O}\right\rangle =0,$ \emph{one can prove an H-theorem only for the
total 0-th order mixing }$S_{0}(t).$ This is done by assuming that $\phi$ is
of the form given in (\ref{ansatz}) so that $\ln\phi$ can be expressed as a
power series in $\sigma$ in which only the constant and linear terms may
depend upon $(x,y),$ i.e., $\ln\phi=\chi(x,y,t)\sigma+\sum_{i=2}\mu
_{i}(t)\sigma^{i}-\ln Z(\chi,t).$ In such a case the term containing
$\overline{O}$ in (\ref{A}) vanishes since by construction $\left\langle
\overline{O}\right\rangle =\left\langle \sigma\overline{O}\right\rangle =0,$
confer (19), and one gets that%
\begin{equation}
\frac{dS_{0}(t)}{dt}=\frac{\nu}{A}\int\!d\sigma\int dx\,dy\,\frac{1}{\phi
}\left\vert \nabla\phi\right\vert ^{2}\geq0.\label{dSo/dt}%
\end{equation}
It is for this reason that the only mixing measure considered in the body of
this paper is the 0-th order $S_{0}(t).$ Non-extensive quantities like
$s_{r}(\sigma,t)$ with $r\neq0$ have been considered as possible
generalizations of the Boltzmann entropy, see e.g.\cite{tsalo} and in the
context of 2D fluid motion\cite{bogho} , confer also reference \cite{branChav}
in which some useful comments are given.

\section{\label{appndC}}

In order to determine the family of probability densities $\phi_{LB}%
(\sigma,x,y,t)$ that reach the lower bound in (\ref{lowerBound}), i.e.,
$\left\langle \left\vert \nabla\ln\phi_{LB}\right\vert ^{2}\right\rangle
=K_{2}^{-1}\left\vert \nabla\omega\right\vert ^{2},$ we write
\[
\ln\phi=:\sum_{n=0}b_{n}(x,y,t)\sigma^{n},
\]
and notice that, due to the normalization condition (\ref{normalztn}), one has
that%
\[
\exp\left(  -b_{0}(x,y,t)\right)  =\int\!d\sigma\,\exp\left(  \sum_{n=1}%
b_{n}(x,y,t)\sigma^{n}\right)  .
\]
It follows then that,
\[
\nabla b_{0}(x,y,t)=-\sum_{n=1}\left\langle \sigma^{n}\right\rangle \nabla
b_{n}(x,y,t),
\]
and consequently that%
\begin{align*}
\nabla\ln\phi & =\sum_{n=1}\left[  \sigma^{n}-\left\langle \sigma
^{n}\right\rangle \right]  \nabla b_{n}(x,y,t),\\
\mathrm{and}\ \ \nabla\omega & =\sum_{n=1}\left\langle \left(  \sigma
-\omega\right)  \left[  \sigma^{n}-\left\langle \sigma^{n}\right\rangle
\right]  \right\rangle \nabla b_{n}(x,y,t).
\end{align*}
It is now easy to see that if for $n\geq2$ one has $\nabla b_{n}=0$ then
$\left\langle \left\vert \nabla\ln\phi\right\vert ^{2}\right\rangle
=\left\langle \left(  \sigma-\omega\right)  ^{2}\right\rangle \left\vert
\nabla b_{1}\right\vert ^{2}$ and also $\nabla\omega=\left\langle \left(
\sigma-\omega\right)  ^{2}\right\rangle \nabla b_{1}(x,y,t),$ i.e.,
$\left\vert \nabla\omega\right\vert ^{2}=K_{2}^{2}\left\vert \nabla
b_{1}\right\vert ^{2}.$ Therefore, in this special case with $\nabla b_{n}=0$
for $n\geq2$ the equality $\left\vert \nabla\omega\right\vert ^{2}%
=K_{2}\left\langle \left\vert \nabla\ln\phi\right\vert ^{2}\right\rangle $
holds. We have shown then that $\phi_{LB}(\sigma,x,y,t) $ can be identified
with $\phi^{\ast}(\sigma,x,y,t)$ as given in Equation (\ref{ansatz}) with
\begin{align}
\chi(x,y,t)  & :=b_{1}(x,y,t),\nonumber\\
\mathrm{and}\ \ \ \mu(\sigma,t)  & :=\sum_{n=2}\mu_{n}(t)\sigma^{n}%
,\ \ \ \nabla\mu_{n}=0,\nonumber
\end{align}
and that it attains the lower bound in (\ref{lowerBound}), i.e.,%
\[
\left\langle \left\vert \nabla\ln\phi_{LB}\right\vert ^{2}\right\rangle
=\frac{\left\vert \nabla\omega\right\vert ^{2}}{K_{2}}.
\]

\section{First and second variation of $\int\!dx\,dy\,\left\langle \left\vert
\nabla\ln\phi\right\vert ^{2}\right\rangle $\label{appendD}}

In this Appendix we investigate the extrema of $\int\!dx\,dy\,\left\langle
\left\vert \nabla\ln\phi\right\vert ^{2}\right\rangle $ when $\phi$\ satisfies
(\ref{normalztn}), (\ref{vortDef}) as well as%
\[
\int d\sigma\,\sigma^{2}\phi(\sigma,x,y,t)=m_{2}(x,y,t).
\]
The quantity we have to vary, let us call it $T(\phi),$ is then%
\[
T(\phi):=\frac{1}{\phi}\left\vert \nabla\phi\right\vert ^{2}+\sum_{n=0}%
^{2}\lambda_{n}(x,y,t)\sigma^{n}\phi,
\]
where $\lambda_{n}(x,y,t)$ are the Langrange multipliers corresponding to the
local-moments constraints $m_{0}(x,y,t)\equiv1,\ m_{1}(x,y,t)=\omega(x,y,t)$
and $m_{2}(x,y,t)$. Up to second-order terms in $\delta\phi$ we have%
\begin{align*}
T(\phi+\delta\phi)-T(\phi)  & =2\nabla\cdot\left(  \delta\phi\nabla\ln
\phi\right)  +\delta_{1}T+\delta_{2}T,\\
\mathrm{with}\ \ \delta_{1}T  & :=\left[  \frac{1}{\phi^{2}}\left\vert
\nabla\phi\right\vert ^{2}-2\frac{\Delta\phi}{\phi}+\sum_{n=0}^{2}\lambda
_{n}(x,y)\sigma^{n}\right]  \delta\phi,\\
\mathrm{and}\ \ \delta_{2}T  & :=\frac{1}{\phi}\left[  \nabla\delta\phi
-\frac{\nabla\phi}{\phi}\delta\phi\right]  ^{2}.
\end{align*}
The first term is a total divergence that, upon integration over $(x,y),$
leads to a vanishing boundary term. The extrema are determined by $\delta
_{1}T=0$ and since $\delta_{2}T\geq0$ it follows that all extrema are minima
of $\int\!dx\,dy\,\left\langle \left\vert \nabla\ln\phi\right\vert
^{2}\right\rangle $ under the given constraints. In order to solve $\delta
_{1}T=0$ it is convenient to introduce $e(\sigma,x,y,t):=\phi^{1/2}$ and
noticing then that%
\[
\Delta e=\frac{1}{2}\left(  \frac{\Delta\phi}{\phi}-\frac{1}{2}\frac
{\left\vert \nabla\phi\right\vert ^{2}}{\phi^{2}}\right)  e,
\]
one sees that $\delta_{1}T=0$ is equivalent to%
\[
\Delta e=\frac{1}{4}\left(  \sum_{n=0}^{2}\lambda_{n}(x,y)\sigma^{n}\right)
e.
\]
Let us denote the minimizer of $T(\phi)$ by $\phi_{ext}(\sigma,x,y,t),$ i.e.,
$\delta_{1}T(\phi_{ext})=0.$ As one can check, this minimizer can be written
as%
\begin{align*}
\phi_{ext}(\sigma,x,y,t)  & =\exp\left[  \mu(\sigma,t)+b_{0}(x,y,t)+\chi
(x,y,t)\sigma\right]  ,\\
\mathrm{where}\ \ \exp\left[  -b_{0}(x,y,t)\right]   & :=%
%TCIMACRO{\dint }%
%BeginExpansion
{\displaystyle\int}
%EndExpansion
\!\!d\sigma\,\exp\left[  \mu(\sigma,t)+\chi(x,y,t)\sigma\right]  ,
\end{align*}
which can thus be identified with $\phi^{\ast}(\sigma,x,y,t)$ as given in
Equation (\ref{ansatz}). In this extremal case, the Lagrange multipliers are,%
\begin{align*}
\lambda_{0}  & =\omega^{2}\left\vert \nabla\chi\right\vert ^{2}-2\omega
\Delta\chi-2\nabla\chi\cdot\nabla\omega\\
& =\left[  K_{2}\right]  ^{-2}\left[  \left(  \omega^{2}-2K_{2}\right)
\left\vert \nabla\omega\right\vert ^{2}-2K_{2}\omega\left(  \Delta
\omega-\nabla\omega\cdot\nabla\ln K_{2}\right)  \right]  ,\\
\lambda_{1}  & =4\Delta\chi-2\omega\left\vert \nabla\chi\right\vert ^{2}\\
& =\left[  K_{2}\right]  ^{-2}\left[  2K_{2}\left(  \Delta\omega-\nabla
\omega\cdot\nabla\ln K_{2}\right)  -2\omega\left\vert \nabla\omega\right\vert
^{2}\right]  ,\\
\lambda_{2}  & =\left\vert \nabla\chi\right\vert ^{2}=\left[  K_{2}\right]
^{-2}\left\vert \nabla\omega\right\vert ^{2},\\
\lambda_{n}(x,y,t)  & =0,\ \ \mathrm{for}\ \ n=3,4,....
\end{align*}

\section{Cases associated with finite-order differental
equations\label{appendixE}}

The differential equations (\ref{omegaN}) and (\ref{linearZ}) are of infinite
order, however, in some cases they reduce to finite--order differential
equations in a non-trivial way. For example, when $\mu(\sigma) $ is such that%
\begin{equation}
\frac{d\mu}{d\sigma}=-2q^{2}\frac{P_{N}(q\sigma)}{Q_{M}(q\sigma)}%
\sigma,\label{rationalDmu}%
\end{equation}
where $P_{N}(x)$ and $Q_{M}(x)$ are polynomials in $x$ of order $N$ and $M,$
respectively, $\lim_{x\rightarrow0}P_{N}(x)=\lim_{x\rightarrow0}Q_{M}(x)$\ and
$q$ is a parameter such that $q\sigma$ is dimensionless. Then equation
(\ref{Zparts}) reads%
\[
2q^{2}\left\langle \frac{P_{N}(q\sigma)}{Q_{M}(q\sigma)}\sigma\right\rangle
=\chi,
\]
and using the recursion operator $\mathcal{L}$ defined in (\ref{opertL}), it
follows that%
\[
2q^{2}P_{N}(q\mathcal{L})\widetilde{\Omega}(\chi)=Q_{M}(q\mathcal{L})\chi.
\]
The remarkable disappearance of the $\sigma-$average can be traced back to the
special $\chi-$dependence of the extremal distributions, confer equation
(\ref{opertL}). Making now use of equation (\ref{Ln}), the linear differential
equation (\ref{linearZ}) that $Z(\chi)$ must satisfy when (\ref{rationalDmu})
holds follows, namely%
\begin{equation}
2q^{2}P_{N}(q\frac{d\ }{d\chi})\frac{dZ}{d\chi}=Q_{M}(q\frac{d\ }{d\chi
})\left(  \chi Z\right)  .\label{finiteODE}%
\end{equation}
In Subsection \ref{finiteNRconservatn}\ref{2nd&4thMomntConservtn} a case with
$N=2$ and $M=0$ was considered. An example with $P_{N}(x)=1$ is provided by a
$\mu(\sigma)$ of the following form,%
\[
\mu_{d}(\sigma):=\left\{
\begin{array}
[c]{c}%
d^{-2}\ln\left[  1-d^{2}q^{2}\sigma^{2}\right]  \ \ \ \mathrm{for}%
\ \ \sigma^{2}<\left[  dq\right]  ^{-2},\\
-\infty\ \ \ \ \mathrm{for}\ \ \ \ \sigma^{2}\geq\left[  dq\right]  ^{-2},
\end{array}
\right.
\]
where $d$ is a pure number. For finite $d$ the vorticity distribution has a
finite support, $\sigma^{2}<\left[  dq\right]  ^{-2},$ while in the limit
$d\rightarrow0$ it approaches a Gaussian. Since%
\[
\frac{d\mu_{d}}{d\sigma}=-\frac{2q^{2}\sigma}{1-d^{2}q^{2}\sigma^{2}},
\]
the differential equation satisfied by the corresponding partition function
$Z_{d}$ is%
\begin{align*}
2q^{2}\frac{dZ_{d}}{d\chi}  & =\left[  1-d^{2}q^{2}\frac{d^{2}\ }{d\chi^{2}%
}\right]  \left(  \chi Z_{d}\right)  ,\\
\mathrm{i.e.,}\ \ \ \frac{d^{2}Z_{d}}{d\chi^{2}}+\frac{2}{\chi}(1+d^{-2}%
)\frac{dZ_{d}}{d\chi}-\frac{Z_{d}}{q^{2}d^{2}}  & =0.
\end{align*}
This is a modified Bessel equation of fractional order. The corresponding
boundary conditions are $Z_{d}(0)=1$ and $Z_{d}^{\prime}(0)=0$. The solution
is:%
\[
Z_{d}(\chi)=\left(  \frac{q}{\chi}\right)  ^{\alpha}Y_{\alpha}(\frac{\chi}%
{qd}),
\]
where $Y_{\alpha}(\chi/qd)$\ is a modified Bessel function of fractional order
and $\alpha:=(d^{2}+2)/2d^{2}.$ Indeed, since in the present case $\exp\mu
_{d}(\sigma)=\left[  1-d^{2}q^{2}\sigma^{2}\right]  ^{1/d^{2}}$\ for
$\sigma^{2}<1/\left(  q^{2}d^{2}\right)  $ and $0$\ otherwise, we have that%
\begin{align*}
Z_{d}(\chi)  & =\int_{-1/qd}^{1/qd}\!d\sigma\,\left[  1-d^{2}q^{2}\sigma
^{2}\right]  ^{1/d^{2}}\exp\chi\sigma\\
& =\frac{1}{qd}\int_{-1}^{1}\!ds\,\left[  1-s^{2}\right]  ^{1/d^{2}}%
\exp\left(  \frac{\chi s}{qd}\right) \\
& \propto\left(  \frac{qd}{\chi}\right)  ^{\alpha}\mathrm{Y}_{\alpha}%
(\frac{\chi}{qd}),
\end{align*}
corresponding to the integral representation of the modified Bessel functions.
As one can check $\lim_{\chi\rightarrow\infty}\widetilde{\Omega}(\chi)=\left[
qd\right]  ^{-1}$and $\lim_{\chi\rightarrow0}d\widetilde{\Omega}%
/d\chi=1/q(3d^{2}+2).$The graph in Fig. 2 shows the corresponding
$\widetilde{\Omega}(\chi)$ for the cases $d=1$ and $d=3.$
%TCIMACRO{\FRAME{ftbpFU}{4.7556in}{6.7948in}{0pt}{\Qcb{Plot of $q\widetilde
%{\Omega}$ as a function of $\chi/q$ for $d=1$ (crosses) and $\ d=3$ (full
%line).}}{\Qlb{fig2}}{hansappd.eps.eps}{\special{ language "Scientific Word";
%type "GRAPHIC";  maintain-aspect-ratio TRUE;  display "USEDEF";
%valid_file "F";  width 4.7556in;  height 6.7948in;  depth 0pt;
%original-width 6.4913in;  original-height 9.2933in;  cropleft "0";
%filename '../../Program Files/Maple 8/MyMaple/hansAppD.eps.eps';file-properties "XNPEU";}%
%}}%
%BeginExpansion
\begin{figure}
[ptb]
\begin{center}
\includegraphics[
height=6.7948in,
width=4.7556in
]%
{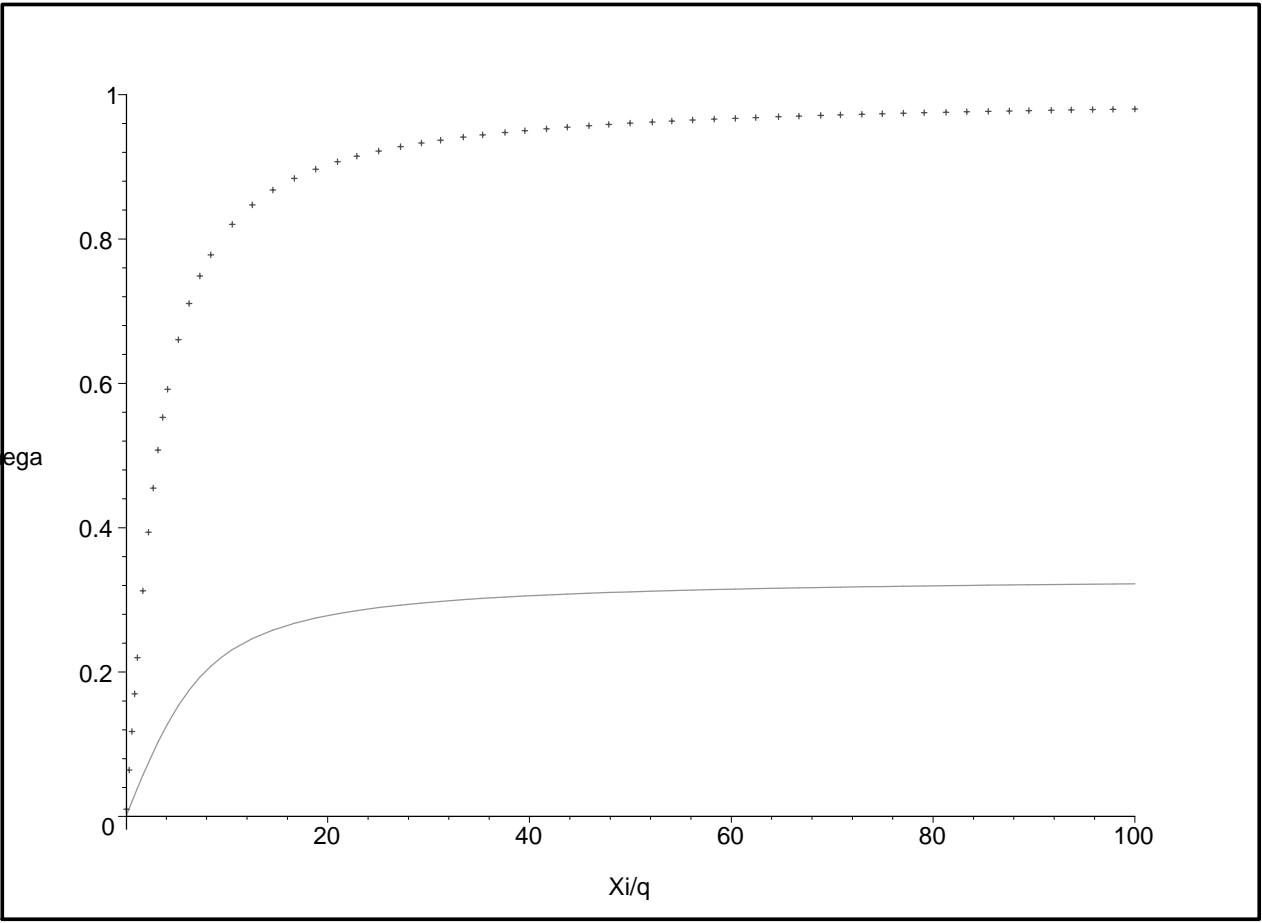}%
\caption{Plot of $q\widetilde{\Omega}$ as a function of $\chi/q$ for $d=1$
(crosses) and $\ d=3$ (full line).}%
\label{fig2}%
\end{center}
\end{figure}
%EndExpansion
\newpage\label{bibliography}\label{bibliografia}

\newpage


\begin{thebibliography}{99}                                                                                               %
\bibitem {arndt}C. Arndt, Information measures, Spinger, Berlin, 2001.

\bibitem {bogho}B. M. Boghosian, \textquotedblleft Thermodynamic description
of the relaxation of 2D turbulence using Tsallis statistics\textquotedblright,
Phys. Rev. E \textbf{53}, 4754 (1996).

\bibitem {brands}H. Brands, J. Stulemeyer, R.A. Pasmanter and T.J. Schep,
\textquotedblleft A mean field prediction of the asymptotic state of decaying
2D turbulence\textquotedblright, Phys. Fluids \textbf{9} 2815, (1998).

\bibitem {branChav}H. Brands, P.-H. Chavanis, R.A. Pasmanter and J. Sommeria,
\textquotedblleft Maximum entropy versus minimum enstrophy
vortices\textquotedblright, Phys. Fluids \textbf{11}, 3465 (1999).

\bibitem {respo}H. Brands, J. Stulemeyer, R.A. Pasmanter and T.J. Schep,
Response to \textquotedblleft Comment on\ `A mean field prediction of the
asymptotic state of decaying 2D turbulence' \textquotedblright, Phys. Fluids
\textbf{10} 1238, (1998).

\bibitem {Capel}H.W. Capel and R.A. Pasmanter, \textquotedblleft Evolution of
the vorticity-area density during the formation of coherent structures in
two-dimensional flows\textquotedblright, Phys. Fluids \textbf{12,} 2514 (2000).

\bibitem {cardoso}O. Cardoso, D. Marteau and P. Tabeling, \textquotedblleft
Quantitative experimental study of the free decay of quasi-two-dimensional
turbulence\textquotedblright, Phys. Rev. E \textbf{49,} 454 (1994)

\bibitem {chavan}P.-H. Chavanis and J. Sommeria, \textquotedblleft
Classification of robust isolated vortices in two-dimensional
turbulence\textquotedblright, J. Fluid Mechanics \textbf{356,} 259--296 (1998).

\bibitem {chavanReview}P.-H. Chavanis, \textquotedblleft Statistical mechanics
of two-dimensional vortices and stellar systems\textquotedblright, in
\textquotedblleft Dynamics and Thermodynamics of Systems with Long Range
Interactions\textquotedblright, T. Dauxois, S. Ruffo, E. Arimondo, M. Wilkens
Eds., Lecture Notes in Physics Vol. \textbf{602}, Springer (2002).

\bibitem {eyink}G. L. Eyink, \textquotedblleft Dissipation in turbulent
solutions of 2D Euler equations\textquotedblright, Nonlinearity \textbf{14},
787 (2001).

\bibitem {flor}G.J.F. van Heijst and J.B. Flor, ``Dipole formations and
collisions in a stratified fluid'', Nature \textbf{340}, 212 (1989)

\bibitem {come}W.H. Matthaeus and D. Montgomery, Comment on \textquotedblleft
A mean field prediction of the asymptotic state of decaying 2D
turbulence\textquotedblright, Physics of Fluids \textbf{10}, 1237 (1998).

\bibitem {lyndenBell}D. Lynden-Bell, \textquotedblleft Statistical mechanics
of violent relaxation in stellar systems\textquotedblright, Mon. Not. R.
Astro. Soc. \textbf{136,} 101 (1967).

\bibitem {marteau}D. Marteau, O. Cardoso, and P. Tabeling, \textquotedblleft
Equilibrium states of 2D turbulence: an experimental study\textquotedblright.
Phys. Rev. E \textbf{51}, 5124-5127 (1995).\ 

\bibitem {matthaeSelect}W.H. Matthaeus, W. T. Stribling, D. Martinez, S.
Oughton and D. Montgomery, \textquotedblleft Selective decay and coherent
vortices in two-dimensional incompressible turbulence\textquotedblright, Phys.
Rev. Letters 66, 2731 (1991) and the references therein.

\bibitem {millerPRL}J. Miller, \textquotedblleft Statistical mechanics of
Euler's equation in two dimensions\textquotedblright, Phys. Rev. Lett.
\textbf{65}, 2137 (1990).

\bibitem {millerPRA}J. Miller, P.B. Weichman, and M.C. Cross,
\textquotedblleft Statistical mechanics, Euler's equation, and Jupiter's Red
Spot\textquotedblright, Phys. Rev. A \textbf{45}, 2328 (1992).

\bibitem {montPFA2}D. Montgomery, X. Shan, and W. Matthaeus, \textquotedblleft
Navier-Stokes relaxation to sinh-Poisson states at finite Reynolds
numbers\textquotedblright, Phys. Fluids A \textbf{5,} 2207 (1993).

\bibitem {onsager}L. Onsager, \textquotedblleft Statistical
Hydrodynamics\textquotedblright, Nuovo Cimento Suppl. \textbf{6}, 279-287 (1949).

\bibitem {ottino}J. M. Ottino, \textquotedblleft Mixing, chaotic advection and
turbulence\textquotedblright, Annual Rev. Fluid Mech.\textbf{\ 22}, 207 (1990).

\bibitem {robertfrans}R. Robert, \textquotedblleft Etat d'\'{e}quilibre
statistique pour l'\'{e}coulement bidimensionnel d'un fluide parfait, C. R.
Acad. Sci. Paris \textbf{311} (S\'{e}rie I), 575 (1990).

\bibitem {robertJSP}R. Robert, ``Maximum entropy principle for two-dimensional
Euler equations'', J. Stat. Phys. \textbf{65}, 531 (1991).

\bibitem {robrelax}R.Robert and J. Sommeria, \textquotedblleft Relaxation
towards a statistical equilibrium state in two-dimensional perfect fluid
dynamics\textquotedblright, Phys. Rev. Lett. \textbf{69}, 2776 (1992).

\bibitem {robsom}R. Robert and J. Sommeria, \textquotedblleft Statistical
equilibrium states for two dimensional flows\textquotedblright, J. Fluid Mech.
\textbf{229}, 291 (1991).

\bibitem {segre}E. Segre and S. Kida, \textquotedblleft Late states of
incompressible 2D decaying vorticity fields\textquotedblright, Fluid Dyn. Res.
\textbf{23}, 89 (1998).

\bibitem {shohat}J. A. Shohat and J. D. Tamarkin, \textsl{The problem of
moments}, Amer. Math. Soc. Math. Surveys, no. 1, 1943.

\bibitem {sommeria}J. Sommeria, C. Staquet, and R. Robert, \textquotedblleft
Final equilibrium state of a two dimensional shear layer\textquotedblright, J.
Fluid Mech. \textbf{233}, 661 (1991).

\bibitem {tsalo}C. Tsallis, \textquotedblleft Possible generalization of
Boltzmann-Gibbs statistics", J. Stat. Phys. \textbf{52}, 479 (1988); See also http://tsallis.cat.cbpf.br/biblio.htm.

\bibitem {mcwilliams}J.C. McWilliams, \textquotedblleft The emergence of
isolated coherent vortices in turbulent flow\textquotedblright, J. Fluid Mech.
\textbf{146}, 21 (1984).
\end{thebibliography}
\end{document}